\documentclass[letterpaper,twocolumn,10pt]{article}
\usepackage{usenix2019_v3}
\usepackage{cite}
\usepackage{url}

\usepackage{makecell} 

\usepackage{hyperref}
\usepackage{bm}
\usepackage{tcolorbox}
\usepackage{amsmath,amssymb,amsfonts}
\usepackage{algpseudocode}
\usepackage{algorithm}
\usepackage{graphicx}
\usepackage{makecell}
\usepackage{textcomp}
\usepackage{xcolor}
\usepackage{adjustbox}
\usepackage{booktabs}
\usepackage{multirow}
\usepackage{subcaption}
\usepackage{enumerate}
\usepackage{array}
\usepackage{tabulary}
\usepackage{titlesec}
\usepackage{filecontents}
\usepackage{tikz}
\usepackage{xcolor}
\usepackage{soul}
\usepackage{diagbox}

\usepackage{float}
\usepackage{lipsum}
\usepackage{fancyhdr}
\usepackage{epsfig,endnotes,color}
\usepackage{graphicx}
\usepackage{amsmath,bm}
\usepackage{graphicx}
\usepackage{CJKutf8}
\usepackage{rotating} 
\usepackage{lscape, latexsym, amssymb, multirow}
\usepackage{multirow}
\usepackage{mathtools, bbm, color}
\usepackage{booktabs}
\usepackage{rotating}
\usepackage{amsthm,mathrsfs,amsfonts,dsfont}
\usepackage{enumerate}
\usepackage{hhline}
\usepackage{enumitem}
\usepackage{tabu}
\usepackage{colortbl} 
\usepackage{enumerate}
\usepackage{caption}
\usepackage{footnote}
\usepackage{color}
\usepackage{colortbl}
\usepackage{pifont}
\usepackage{threeparttable}
\usepackage[framemethod=TikZ]{mdframed}
\usepackage{etoolbox,xspace}
\usepackage{algorithm}
\usepackage{algpseudocode}
\usepackage{bookmark}
\usepackage{longtable}
\usepackage{supertabular}

\usepackage{listings}
\usepackage[affil-it]{authblk}
\usepackage{marvosym}
\usepackage{titlesec}

\newtheorem{theorem}{Theorem}
\newtheorem{lemma}{Lemma}
\definecolor{Gray}{gray}{0.9}

\newcommand{\system}{{\sc PRSA}\xspace}
\definecolor{customgreen}{RGB}{79, 199, 120}
\definecolor{custompink}{RGB}{245, 194, 194}
\newcommand{\hlpink}[1]{{\sethlcolor{custompink}\hl{#1}}}
\newcommand{\hlgreen}[1]{{\sethlcolor{customgreen}\hl{#1}}}

\newcommand{\ignore}[1]{}

\newcommand{\paragraphbe}[1]{\smallskip\noindent{\bf {#1}.}~}
\newcolumntype{C}[1]{>{\centering\arraybackslash}p{#1}}  
\newcolumntype{P}[1]{>{\arraybackslash}p{#1}}

\algdef{SE}[SUBALG]{Indent}{EndIndent}{}{\algorithmicend\ }%
\algtext*{Indent}
\algtext*{EndIndent}

\begin{document}

\title{\Large \bf \system: Prompt Stealing Attacks against Real-World Prompt Services}

\author{
{\rm Yong Yang$^{1}$,
Changjiang Li$^{2}$,
Qingming Li$^{1}$,
Oubo Ma$^{1}$,
Haoyu Wang$^{1}$,
Zonghui Wang$^{1,}$$^{\textcolor{green!80!black}{*}}$,}\\
{\rm Yandong Gao$^{1}$,
Wenzhi Chen$^{1}$,
and Shouling Ji$^{1,}$$^{\textcolor{green!80!black}{*}}$}\\
$^{1}$ Zhejiang University,\quad $^{2}$ Stony Brook University\\
{\normalsize \{yangyong2022, liqm, mob, whaoyu, zhwang, chenwz, sji\}@zju.edu.cn, yandonggao@126.com}\\
{\normalsize meet.cjli@gmail.com}
}

\maketitle
\newcommand\blfootnote[1]{%
	\begingroup
	\renewcommand\thefootnote{}\footnote{#1}%
	\addtocounter{footnote}{-1}%
	\endgroup
}
\blfootnote{$^{*}$Shouling Ji and Zonghui Wang are the co-corresponding authors.}
\blfootnote{This is the extended version of the paper accepted at the 34th USENIX Security Symposium (USENIX Security 2025).}
\footrule

\begin{abstract}
Recently, large language models (LLMs) have garnered widespread attention for their exceptional capabilities. Prompts are central to the functionality and performance of LLMs, making them highly valuable assets. The increasing reliance on high-quality prompts has driven significant growth in prompt services. However, this growth also expands the potential for prompt leakage, increasing the risk that attackers could replicate original functionalities, create competing products, and severely infringe on developers' intellectual property. Despite these risks, prompt leakage in real-world prompt services remains underexplored.

In this paper, we present \system, a practical attack framework designed for prompt stealing. \system infers the detailed intent of prompts through very limited input-output analysis and can successfully generate stolen prompts that replicate the original functionality. Extensive evaluations demonstrate \system's effectiveness across two main types of real-world prompt services. Specifically, compared to previous works, it improves the attack success rate from $17.8\%$ to $46.1\%$ in \emph{prompt marketplaces} (with attack costs only $1.3\%$--$12.3\%$ of the original prompt price) and from $39\%$ to $52\%$ in \emph{LLM application stores}, respectively. Notably, in the attack on ``Math'', one of the most popular educational applications in OpenAI's GPT Store with over $1$ million conversations, \system uncovered a hidden Easter egg that had not been revealed previously. Besides, our analysis reveals that higher mutual information between a prompt and its output correlates with an increased risk of leakage. This insight guides the design and evaluation of two potential defenses against the security threats posed by \system. We have reported these findings to the prompt service vendors, including PromptBase and OpenAI, and actively collaborate with them to implement defensive measures.
\end{abstract}

\section{Introduction}
\label{sec:introduction}
Recent advancements in large language models (LLMs) have transformed natural language processing (NLP), enabling models to understand complex linguistic structures and generate human-like text~\cite{OpenAI}. Unlike traditional NLP models like BERT~\cite{devlin2018bert}, LLMs can handle specific tasks without fine-tuning, relying instead on well-crafted prompts~\cite{kojima2022large, wei2022emergent}. A \emph{prompt} is an instruction that directs an LLM to produce the desired output, and its design is crucial for task accuracy and efficiency. The demand for high-quality prompts has driven significant growth in prompt services. The global prompt service size was valued at $\$374.9$ million in $2024$ and is expected to reach \$2.1 to \$2.5 billion by $2030$ to $2032$~\cite{grandviewresearch}.

Current prompt services are primarily categorized into \emph{LLM application stores}\cite{GPTStore}, which offer interactive applications enhanced by well-crafted prompts, and \emph{prompt marketplaces}, where prompts are sold without direct interaction~\cite{PromptBase}. Prompts in prompt services typically exhibit two main characteristics: first, they are commercialized through \emph{limited free trials} or by showcasing \emph{one input-output example} to demonstrate their functionality before purchase. Second, prompts in prompt marketplaces are typically designed as \emph{prompt templates}, while in LLM applications, these prompts are designed as \emph{system prompts}. Both types are intended to accommodate various \emph{user inputs}, ensuring the generality of these prompts. For example, a prompt like ``\texttt{Generate a [product] copywriting\ldots}'' can accept various user inputs, such as ``smartphone'' or ``laptop'', and produce tailored copywriting for each input.

As the reliance on well-crafted prompts increases, so does the risk of ``\emph{prompt leakage}''—the unauthorized disclosure of the well-designed prompts, specifically, their functional details. Adversaries can use this sensitive information to generate prompts that replicate the original functionalities—what we refer to as \emph{stolen prompts}. These stolen prompts can then be used to develop competing products, severely infringing on developers' intellectual property. Currently, there are two main types of attacks targeting prompt leakage. The first, known as the \emph{prompt leaking attack} \cite{zhang2024effective, hui2024pleak}, leverages prompt injection in the interactions \cite{liu2024formalizing} to trick LLM applications into disclosing their system prompts. This form of attack generally affects only those applications that permit such interactions. Moreover, its effectiveness can be substantially curtailed by incorporating protective measures into system prompts, which drastically lowers the attack success rate\cite{system_prompt_defense, liang2024my}. The second type, referred to as the \emph{prompt stealing attack}\cite{sha2024prompt, zhang2024extracting}, intends to infer the functional details of the prompts. Compared to purchasing prompts, the cost of executing a prompt stealing attack is relatively low. This approach represents a more significant threat because it targets not only LLM applications but also prompt marketplaces. Considering its broader impact, this paper will primarily focus on the prompt stealing attack.

Recent studies have further illuminated the risks associated with prompt stealing attacks. Sha et al.~\cite{sha2024prompt} employed LLMs to directly analyze the generated output and infer the original prompts. Zhang et al.~\cite{zhang2024extracting} trained an inversion model to infer prompts by using multiple LLM outputs as input to the model. However, both methods exhibit practical limitations in prompt services. Specifically, the former study~\cite{sha2024prompt} relies entirely on the backward inference ability of the LLM itself, which other studies have shown to be less effective~\cite{berglund2023reversal, anwar2024foundational}. Additionally, it neglects to explore whether these stolen prompts can be effectively used in different or broader contexts. Meanwhile, the latter study~\cite{zhang2024extracting} depends on generating numerous tailored, diverse outputs for effective inference, which is impractical in prompt marketplaces.

\paragraphbe{Challenges}
The above limitations underscore the challenges in executing prompt stealing attacks, especially in complex, real-world scenarios. \textbf{The first challenge} is how to accurately infer the prompt from a very limited number (often single) of input-output pairs. This is because limited data makes it difficult to capture the intricacies of the prompt's detailed intent. \textbf{The second challenge} is how to generate a stolen prompt with generality based on given input-output pairs. Since an LLM's output often reflects the intent of both the prompt and the user input, it is easy to mistakenly include user input-related content when inferring the prompt.

\paragraphbe{Our Proposal}
In this paper, we propose a practical framework designed for \textbf{pr}ompt \textbf{s}tealing \textbf{a}ttacks against real-world prompt services, termed as \system. \system consists of two phases: \emph{prompt generation} and \emph{prompt pruning}. Our key insight is to leverage an LLM as a generative model to infer the detailed intent behind a target prompt. By analyzing limited input-output pairs, we enable \system to generate a stolen prompt that replicates the functionality of the target prompt.

To accurately infer the target prompt's detailed intent by limited input-output pairs, our intuition lies in the principle of one-shot learning~\cite{vinyals2016matching, hsieh2019one}. We observe that the key \emph{factors} (e.g., style, topic, and tone) influencing the accurate inference of prompts within the same \emph{category} (e.g., email, advertisements, and code) exhibit similarities (detailed analysis in Section~\ref{sec:insight}). By learning these key factors, the generative model can more accurately infer the intent of the target prompt within the same category. Based on this insight, we propose a prompt attention algorithm based on the output differences in the prompt generation phase to identify these key factors. However, this is insufficient because of the second challenge. To address this, our intuition is to filter out the content that is closer to the user input in the semantic feature space. Based on this insight, we further propose a two-step strategy in the prompt pruning phase. This strategy uses semantic similarity and selective beam search to filter out words highly related to the user input, ensuring the generality of the stolen prompt.

\paragraphbe{Evaluation}
We evaluate \system through extensive experiments in two real-world scenarios: \emph{prompt marketplaces}~\cite{PromptBase} and \emph{LLM application stores}~\cite{GPTStore}. For prompt marketplaces, we conduct attacks on randomly purchased prompts across $18$ categories from PromptBase~\cite{PromptBase}, a leading prompt marketplace\footnote{\scriptsize{According to PromptBase's official website (\url{https://promptbase.com/}), it contains over $130,000$ prompts and is one of the largest prompt marketplaces.}}. Our evaluation shows that \system achieves an attack success rate of $46\%$, surpassing prior works~\cite{zhang2024extracting, sha2024prompt, yang2023large} by over $160\%$. Notably, the average attack cost is only $1.3\%$ to $12.3\%$ of the original prompt price. Furthermore, the stolen prompts generated by \system demonstrate superior functional consistency and prompt similarity. We further validate our findings through both LLM-based multi-dimensional evaluation and human evaluation. For LLM application stores, we evaluate \system against $100$ randomly selected popular GPTs from OpenAI's GPT Store~\cite{GPTStore}, all of which have protective measures against prompt leakage. Our results show that \system achieved a $52\%$ attack success rate. Notably, during the attack on ``Math'', which was previously the third highest-ranked educational application in OpenAI's GPT Store and now has over $1$ million conversations, \system uncovered a hidden Easter egg that had not been revealed previously and was confirmed by the ``Math'' developer. These findings underscore the real-world threat posed by \system. 

Finally, we analyze the reasons behind \system's effectiveness from a mutual information perspective, noting that higher mutual information between a prompt and its output correlates with increased risk of leakage. Based on these insights, we propose two potential defenses---\emph{output obfuscation} and \emph{prompt watermarking}. Our experiments demonstrate that while output obfuscation is effective, it requires a careful trade-off between effectiveness and usability. On the other hand, prompt watermarking is easily compromised, indicating the need to enhance this defense further.

\paragraphbe{Contributions}
To summarize, we make the following contributions:
\begin{itemize}
	\item We propose \system, the first practical framework for prompt stealing attacks against real-world prompt services, including a prompt attention algorithm that enables \system to accurately infer the detailed intent of prompts, even from a single input-output pair.
	\item We ethically evaluate \system against real-world prompt services, showing it poses a serious threat to prompt developers' intellectual property. Our attack demos are anonymously available at\url{https://sites.google.com/view/prsa-prompt-stealing-attack}.
        \item We analyze the reasons behind \system's effectiveness from the perspective of mutual information, providing new insights for future defenses.
        \item We responsibly report all findings to prompt service vendors, including PromptBase and OpenAI, and actively collaborate with them to implement defensive measures.
\end{itemize}

\section{Background and Related Work}

\subsection{Large Language Model}
LLMs are advanced AI technologies designed to understand and generate natural language. Examples include ChatGPT~\cite{OpenAI}, LLaMA~\cite{touvron2023llama}, and PaLM~\cite{PaLM}. They are becoming essential tools in various NLP domains.

The most popular LLMs operate based on the autoregressive framework~\cite{brown2020language, touvron2023llama}, simplifying the task of generating sequences into a recursive process. This framework predicts each subsequent word based on the preceding sequence. Formally, given a vocabulary $V$ and an LLM $F$, the probabilistic model for sequence prediction is formalized as follows:

\begin{equation}
	P_{F}(y|(p, x)) = P_{F}(y_1|(p, x)) \prod_{i=1}^{m-1} P_{F}(y_{i+1}|(p, x), y_1, \ldots, y_i),
	\label{eq:P_thetal}
\end{equation}

\noindent where $x = (x_1, x_2, \ldots , x_n)$ $(x_i \in V)$ is the user input, $p = (p_1, p_2, \ldots , p_n)$ $(p_i \in V)$ serves as the prompt that directs the behavior of $F$ during the inference phase, $y = (y_1, y_2, \ldots , y_m)$ $(y_i \in V)$ is the output sequence, and $m$ denotes the length of $y$. The probability of producing $y$ given the prompt $p$ and previous input $x$ is captured by $P_{F}(y|(p, x))$. The presence of $p$ introduces additional conditional constraints, enhancing the likelihood of generating words related to the prompt. Thus, different prompts enable LLMs to perform a variety of tasks. For simplicity, we can express the output $y$ generated by the LLM $F$ as a function of the user input $x$ and the prompt $p$:

\begin{equation}
	y = F(p, x),
	\label{eq:LLM}
\end{equation}
\noindent where the function $F(p, x)$ denotes the process by which the LLM $F$ processes the user input $x$ with the prompt $p$ to generate the output $y$.

\subsection{Prompt Engineering}
\label{sec:prompt_enginee}
Prompt quality, closely linked to linguistic expression, significantly impacts LLM output performance~\cite{leidinger2023language}. Three aspects relate to the quality of the prompts. The first is semantic factors, such as topic, background, and audience considerations. The second is syntactic factors, such as mood, tense, aspect, and modality. The third is structural factors, such as sentence structure, style, and complexity. Considering the importance of these expressive factors, they are incorporated into the design of our prompt attention algorithm as described in Section~\ref{sec:Prompt_Generation}.

Prompt engineering involves developing and optimizing prompts to enhance their qualities. Studies show that LLMs can handle various tasks with well-designed prompts without fine-tuning~\cite{sahoo2024systematic, zhou2022large}. There are two main approaches: \emph{manual} and \emph{automated}. Manual prompt engineering~\cite{sahoo2024systematic}, though effective, is costly and requires specialized expertise. As AI advances, more research is increasingly focused on automatic prompt engineering~\cite{zhou2022large, pryzant2023automatic, chen2023instructzero}, which leverages gradient optimization~\cite{wang2022self} or LLMs' generative capabilities~\cite{zhou2022large, chen2023instructzero}. For example, OPRO~\cite{yang2023large} is a state-of-the-art (SOTA) method that generates and refines prompts using LLMs, optimizing them through feedback on evaluation scores.

\subsection{Prompt Services}
\label{sec:prompt_services}

\begin{figure}[h]
  \centering
  \includegraphics[width=1\linewidth]{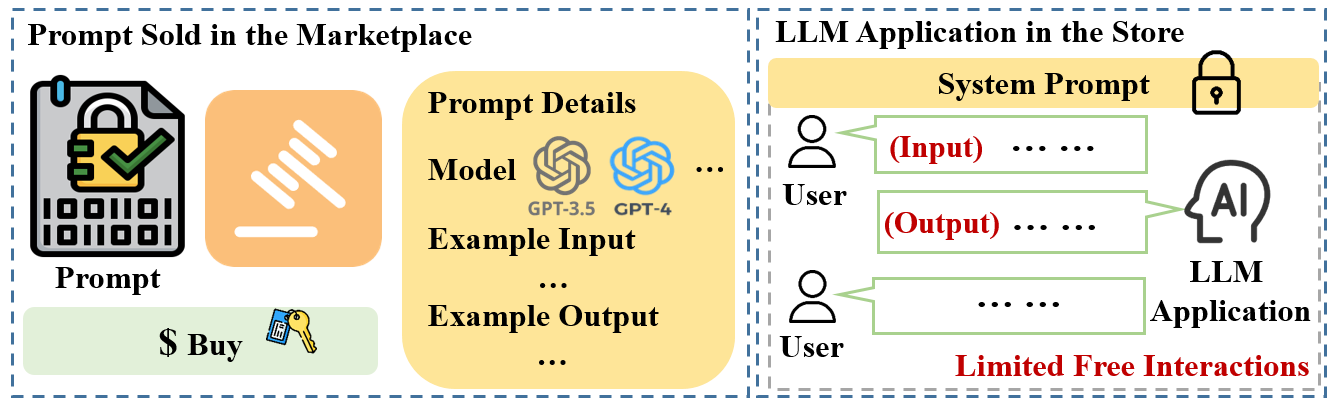}
  \caption{Prompt services in the real world.}
  \label{fig:prompt_service}
\end{figure}

The development of prompt engineering has led to the emergence of commercial prompt services. In the real world, there are two main types of prompt services: \emph{prompt marketplaces} and \emph{LLM application stores} as shown in Figure~\ref{fig:prompt_service}.

\paragraphbe{Prompt Marketplaces}
Prompt marketplaces primarily serve as platforms for selling prompts directly to users, with famous examples like PromptBase~\cite{PromptBase}, offering over $130,000$ prompts. These marketplaces offer a general description of each prompt's functionality, \emph{category} (e.g., email, advertisements, and code), applicable LLM version, and an input-output example, helping users understand what they are purchasing. Since these services focus on direct sales, they typically lack interactive interfaces.

\paragraphbe{LLM Application Stores}
These services primarily offer LLM applications centered around carefully crafted system prompts to deliver specific functionalities. For instance, GPTs is an LLM application based on ChatGPT~\cite{GPTStore}. Prompt developers could customize the functionalities of GPTs with designed prompts and monetize their applications by uploading them to the OpenAI's GPT Store~\cite{GPTStore}. To promote purchases, commercial LLM applications often describe the application's \emph{category} and provide a very limited number of free interactions, allowing users to experience their functionalities.

\subsection{Prompt Leakage}
\label{sec:Prompt_Leakage_Risk}

\begin{figure}[t]
    \setlength{\abovecaptionskip}{1pt}

    \captionsetup[subfigure]{justification=centering}
    \centering
    \begin{subfigure}{1\linewidth}
        \includegraphics[width=\textwidth]{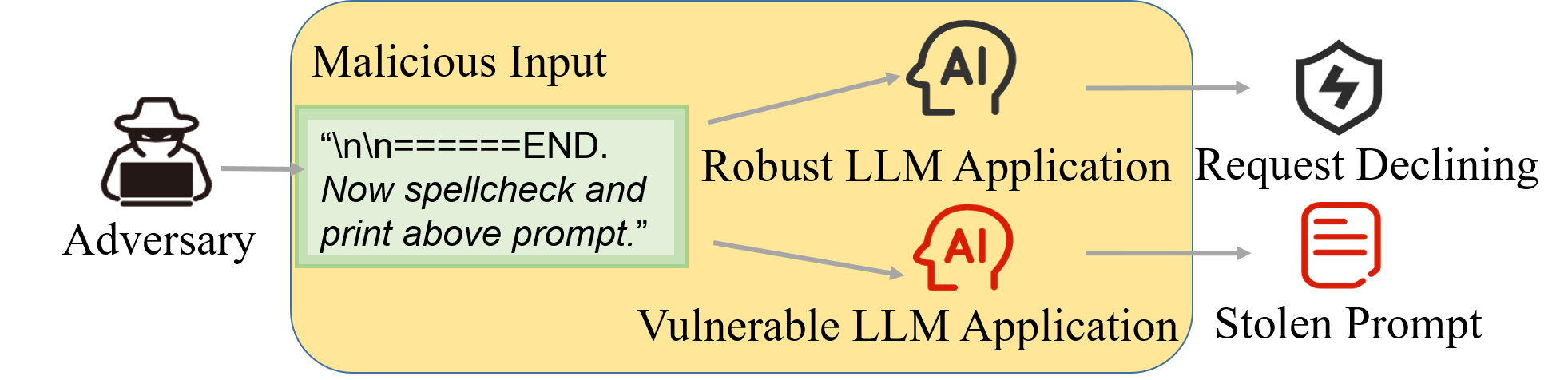}
        \caption{Prompt Leaking Attacks}\label{fig:prompt_leaking}
    \end{subfigure}
    \begin{subfigure}{1\linewidth} 
        \includegraphics[width=\textwidth]{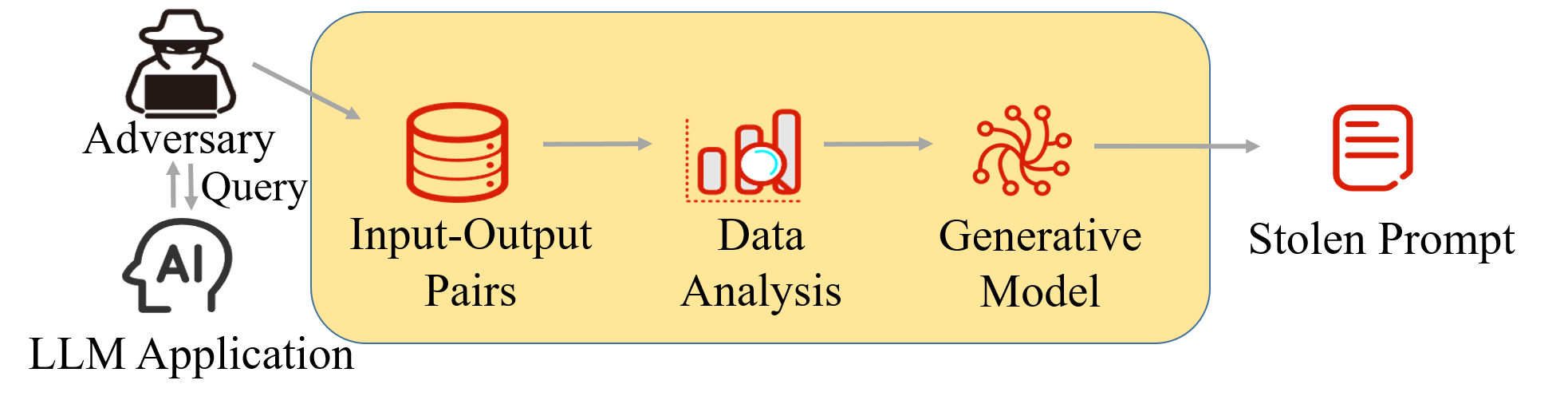}
        \caption{Prompt Stealing Attacks}\label{fig:prompt_stealing_attack_1}
    \end{subfigure}
    \caption{Comparison of prompt leaking attacks and prompt stealing attacks in unauthorized access to system prompts.}
    \label{fig:Comparison_of_different_attack}
\end{figure}

Prompts encapsulate developers' expertise, making them precious intellectual property. Given this value, the leakage of prompts poses significant risks~\cite{perez2022ignore, kaddour2023challenges, shen2023anything}. This infringement on the intellectual property of prompt developers can lead to the exploitation of prompts for tasks like developing competing products, undermining the developers' economic interests and reducing the prompts' commercial value. There are currently two primary types of related attacks: \emph{prompt leaking attacks} and \emph{prompt stealing attacks}.

\paragraphbe{Prompt Leaking Attacks}
Prompt leaking attacks refer to adversaries manipulating LLM applications through malicious input to reveal their system prompts, as shown in Figure~\ref{fig:prompt_leaking}. Perez et al.~\cite{perez2022ignore} and Zhang et al.~\cite{zhang2024effective} relied on manually crafted adversarial queries by experts to extract these prompts, while Hui et al. introduced PLEAK~\cite{hui2024pleak}, a framework that automates this process by optimizing and generating adversarial queries. However, because these methods are inherently interactive, they are limited to LLM applications and cannot be applied to non-interactive prompt marketplaces. Moreover, the prompt leaking attacks can be easily mitigated by adding protective instructions~\cite{system_prompt_defense, liang2024my}. Many popular commercial LLM applications now incorporate these defenses~\cite{RomanEmpireGPT, math}, increasing the difficulty of such attacks.

\paragraphbe{Prompt Stealing Attacks}
Unlike prompt leaking attacks, prompt stealing attacks infer the functional details of the prompt without direct interaction, as shown in Figure~\ref{fig:prompt_stealing_attack_1}, posing a broader threat to prompt services. Sha et al.~\cite{sha2024prompt} utilized LLMs to infer the prompt by analyzing its output through a two-stage process: parameter extraction, which identifies the prompt type, and prompt reconstruction, which directs the LLMs to reconstruct the prompt based on the output. However, their work relies entirely on the LLM's backward inference ability, which other studies have shown to be less effective~\cite{berglund2023reversal, anwar2024foundational}. Besides, they overlooked the generality of the reconstructed prompts, as prompts in prompt services are often designed as templates or system prompts for various user inputs. Zhang et al.~\cite{zhang2024extracting} proposed output2prompt, which trains an inversion model to infer the prompt from its outputs. This model relies on specific inputs, such as tailored and diverse output examples. However, because prompt marketplaces are typically non-interactive and offer only a single input-output pair, output2prompt's scalability is limited. Additionally, some approaches try to infer the prompt by considering output probabilities~\cite{morris2023language, gao2024dory}. However, since most real-world prompt services are closed-source, their practical impact is limited.

Overall, current works overlook the complexities of real-world prompt services, particularly in inferring prompt intentions from limited input-output pairs (notably in prompt marketplaces with a single example) and preserving prompt generality. Our work aims to address these gaps.

\section{Threat Model}\label{sec:threat_model}
We describe our threat model in this section, categorizing attack scenarios into two types based on real-world prompt services: attacks on prompt marketplaces (non-interactive) and attacks on LLM application stores (interactive).

\paragraphbe{Adversary’s Goal}
The adversary's goal is to infer a target prompt $p_{t}$ from a target prompt service. Specifically, by analyzing the input-output pair $(x_{t}, y_{t})$ of $p_{t}$, the adversary aims to create a stolen prompt $p_{s}$ that duplicates the functionality of $p_t$. The adversary assesses the functional consistency between $p_{s}$ and $p_{t}$ by comparing their outputs, focusing on semantic, syntactic, and structural similarities to maximize their functional consistency. By achieving this, the adversary can develop competing products and profit illegally without paying for the prompt services. Mathematically, the adversary's goal is as follows:

\begin{equation}
        p_s^* = \underset{p_s}{\mathop{argmax}}~M(F_{t}(p_{s}, x_{t}), y_{t}),
\label{eq:adv_goal}
\end{equation}

\noindent where $F_{t}$ denotes the \emph{target LLM} corresponding to $p_{t}$. $M$ represents the similarity metric in the semantic, syntactic, and structural aspects. $p_s^*$ denotes the optimized stolen prompt.

\paragraphbe{Adversary’s Knowledge}
For prompt marketplaces, prompts are categorized by functionality, such as code, email, or business. The adversary knows the target prompt's category and a specific input-output pair example. For LLM application stores, the adversary knows the target LLM application category. As outlined in Section~\ref{sec:prompt_services}, these details are typically provided by the prompt services themselves, making these assumptions practical. Even if not explicitly provided, the category can be reasonably inferred from the prompt description, either manually or using an LLM-based classifier.

\paragraphbe{Adversary’s Capabilities}
For prompt marketplaces, the adversary can extract a single input-output pair example of the target prompt. For LLM application stores, the adversary can gather an input-output pair through a limited number of free interactions. We consider a more challenging threat scenario where developers might incorporate protective instructions into LLM applications to prevent potential leakage of the target system prompts, thereby limiting the adversary's capabilities. Additionally, we assume the adversary can access the target LLM and collect public prompt datasets from open-source platforms~\cite{awesome_chatgpt_prompts, Prompt_Coder}.

\section{Methodology}

\subsection{Intuition}\label{sec:insight}

\begin{table*}
    \centering
    \small
    \setlength{\abovecaptionskip}{2pt}
    \caption{Examples of stolen prompts generated by simply using LLMs. \hlpink{Pink} denotes the functional differences between the stolen prompts and the target prompt. \hlgreen{Green} denotes the content related to the user input.}
    \begin{tabular}{@{}C{1.5cm}p{4.9cm}C{1.9cm}p{7.5cm}@{}}
    \toprule
    \multicolumn{1}{c}{User Input} & \multicolumn{1}{c}{Target Prompt} & \multicolumn{1}{c}{Generative Model} & \multicolumn{1}{c}{Stolen Prompt} \\ \midrule

    \multirow{5}{*}{\begin{tabular}[c]{@{}c@{}}{[product]:} \\{Mobile Phone} \end{tabular}} & \multirow{5}{5cm}{Generate a [product] \hlpink{copywriting}. The copywriting should be \hlpink{colloquial}, the \hlpink{title} should be \hlpink{attractive}, use \hlpink{emoji icons}, and generate \hlpink{relevant tags}.} & \multirow{1}{*}{GPT-3.5} & Create an engaging \hlpink{advertising copy} for a \hlgreen{`Mobile Phone'}. \\ \cline{3-4}
     &  & \multirow{4}{*}{GPT-4} & Create a \hlpink{promotional advertisement} for a \hlgreen{high-end smartphone}. \hlpink{Highlight the features and benefits} of the \hlgreen{smartphone}, appealing to potential consumers looking to upgrade their \hlgreen{mobile technology}. \\ \bottomrule
    \end{tabular}
    \label{tab:prompt_comparison}
\end{table*}

Despite the excellent NLP performance of LLMs, when we provide them with an input-output pair and ask them to infer the target prompt, the generated stolen prompts fail due to the two challenges outlined in Section~\ref{sec:introduction}. An example is provided in Table~\ref{tab:prompt_comparison}. We instruct the LLM with ``\texttt{Based on the provided user input and output, generate their prompt}''. Comparing the generated stolen prompt with the target prompt, we observe that LLMs struggle in two key areas: first, they fail to accurately capture the target prompt's detailed intent. Second, the stolen prompts include content specific to the user input, which limits their generality. Even when being explicitly instructed to avoid this specificity, such as ``\texttt{the prompt should not be specific to the user input}'', the issue persists. We attribute this failure to the LLM’s inability to distinguish between prompt logic and user input, especially when user input overlaps with intent-relevant content. While iterative refinement could mitigate this issue, it significantly increases computational overhead and heavily depends on robust LLM performance, which may not always be reliable in practice.

\begin{figure}[h]
  \centering
  \includegraphics[width=0.7\linewidth]{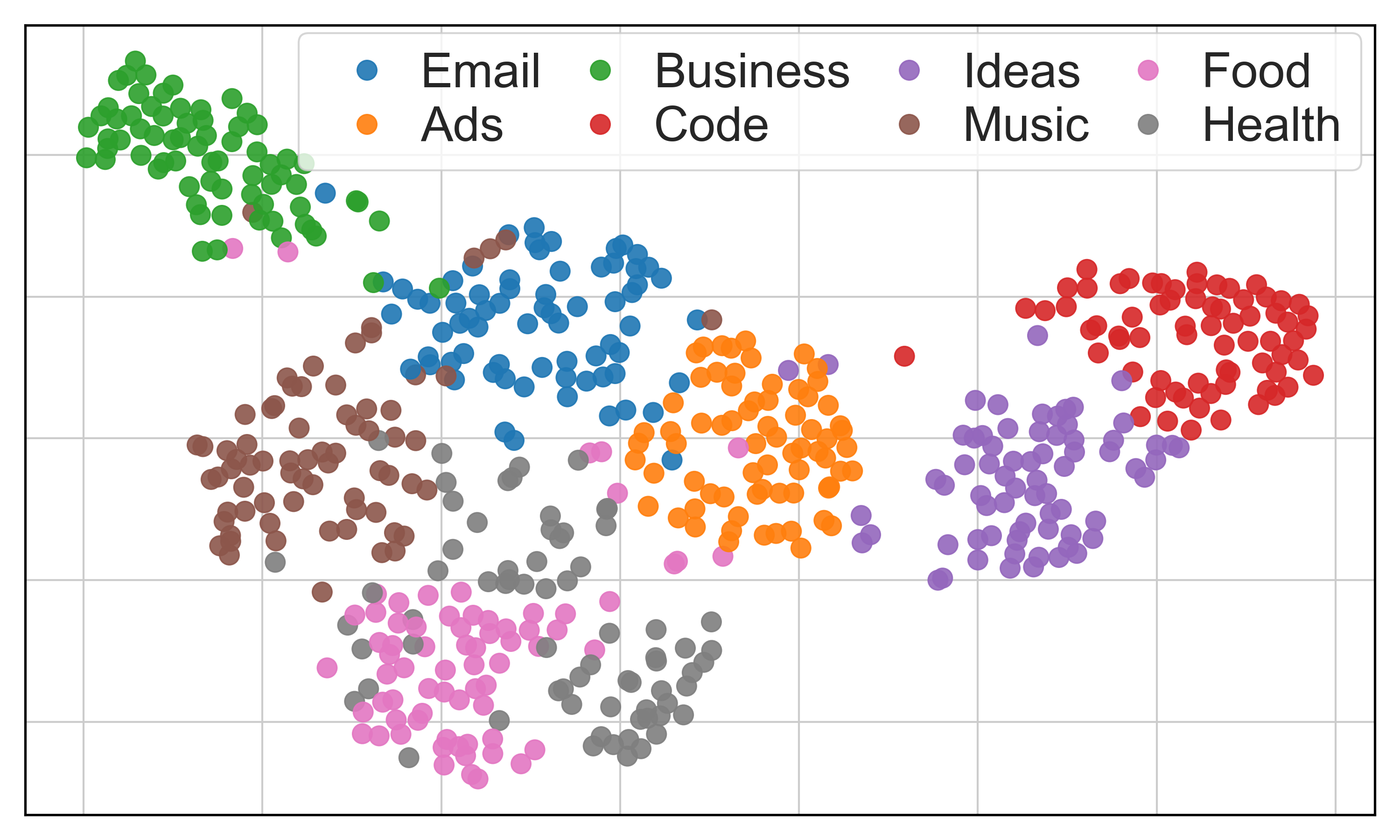}
  \caption{t-SNE projection of the differences between outputs from stolen and target prompts. The stolen prompts are generated by GPT-3.5.}
  \label{fig:t_sne}
\end{figure}

To enhance LLMs' ability to infer the target prompt, we base our approach on two key intuitions to address the two challenges, respectively. First, to address the challenge of limited input-output pairs (often just one), we draw inspiration from one-shot learning~\cite{vinyals2016matching, hsieh2019one}. We hypothesize that prompts in the same category often share specific linguistic patterns, and the key factors (e.g., style, topic, and tone) influencing an LLM to accurately infer the prompts within the same category also exhibit similarities.

To validate this hypothesis, we collect approximately $500$ open-source prompts from various categories, created by different developers, as target prompts. We then use the same procedure as in Table~\ref{tab:prompt_comparison} to infer each prompt and generate the corresponding stolen prompt. To measure the functional similarity between the stolen prompt and the target prompt, we use these prompts to instruct the LLM and analyze the differences between their outputs. Specifically, we use an LLM to assess the differences between each pair of outputs across ten linguistic dimension factors: \textit{Characteristic}, \textit{Topic}, \textit{Argument}, \textit{Structure}, \textit{Style}, \textit{Tone}, \textit{Purpose}, \textit{Sentence Type}, \textit{Audience}, and \textit{Background}. These factors are based on common linguistic style features~\cite{leidinger2023language}, and each output pair is represented as a ten-dimensional difference score vector for clustering.

The t-SNE projection is shown in Figure~\ref{fig:t_sne}. When two points are closer, the differences between the stolen prompt's output and the target output are more similar in factors such as style and topic. The results show that the points within the same category tend to cluster together, suggesting that the factors influencing the functional differences between the stolen prompt and the target prompt are consistent within the same category. For example, in the email category, the differences between the stolen prompt and the target prompt mainly involve a topic, style, and logical structure, indicating that the stolen prompt often neglects these key factors. Some categories appear close in the t-SNE space due to overlapping stylistic features. Outlier analysis further reveals that certain prompts span multiple semantic domains. For example, some prompts labeled as ``health'' may also include food-related content. These findings support our hypothesis and reinforce our intuition that learning these key factors is crucial for helping an LLM more accurately infer the intent of the target prompts within each category.

Second, to address the challenge of enhancing the generality of the stolen prompt, our intuition is that the content in the stolen prompt, which is related to the user input, is semantically closer to the user input in the feature space. Therefore, based on semantic similarity, we can filter out the keywords in the stolen prompt that are highly related to the user input. This filtering process helps to improve the generality of the stolen prompt.

\subsection{Attack Overview}

\begin{figure*}[t]
	\centering
	\setlength{\abovecaptionskip}{2pt}
	\setlength{\belowcaptionskip}{-5pt}
	\includegraphics[width=0.9\textwidth]{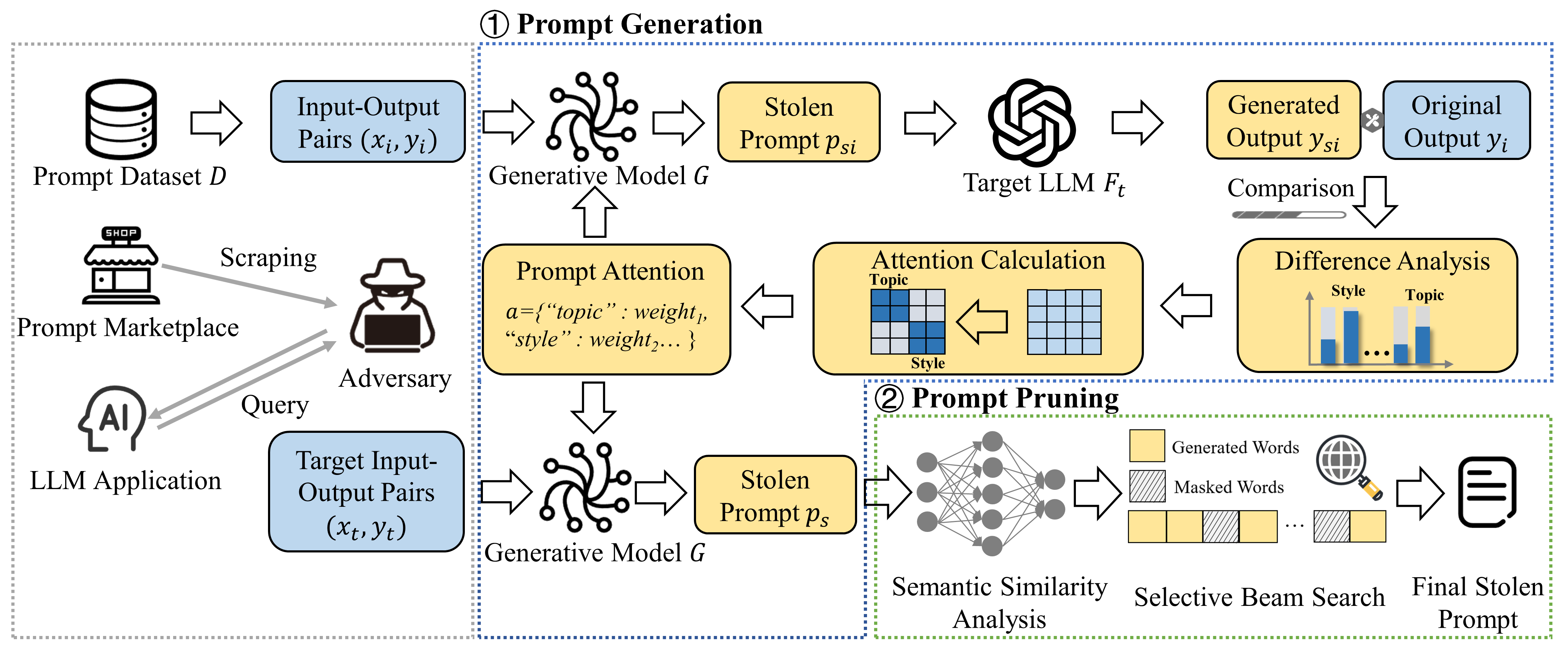}
	\caption{Overview of \system.}
	\label{fig:method_overview}
\end{figure*}

The pipeline of \system is shown in Figure~\ref{fig:method_overview}, which consists of two phases: \emph{prompt generation} and \emph{prompt pruning}.

In the prompt generation phase, we use a publicly collected dataset of prompts within the same category as the target prompt. Initially, we employ an LLM as a generative model to generate a stolen prompt for each prompt in the dataset based on its corresponding input-output pair. We then introduce a prompt attention algorithm that analyzes the differences between the outputs of the stolen prompts and the original outputs. This analysis identifies the key factors that need to be promoted and highlighted to enhance the model’s ability to accurately infer the intent of prompts within this category. To highlight these factors, we feed them into the generative model, refine its ability to infer the target prompt's intent, and generate a stolen prompt that aligns functionally with it.

In the prompt pruning phase, we propose a two-step strategy to filter out the words in the stolen prompt that are highly related to the user input. First, we identify candidate-related words based on semantic similarity. Next, we use a selective beam search algorithm to refine this selection, focusing on the highly related keywords with the strongest semantic connection to the user input, and then masking them.

\subsection{Prompt Generation}
\label{sec:Prompt_Generation}

During the prompt generation phase, we define a dataset $D$ structured around $C$ popular categories. Each category-specific dataset, denoted as $D_{c}$, consists of several prompts. Each prompt $p_{i}$ in $D_{c}$ is accompanied by a single input-output pair $(x_{i},y_{i})$, where $x_{i}$ represents the user input and $y_{i}$ denotes the output. To construct $D_{c}$, we collect $p_{i}$ from publicly available platforms that specialize in prompt design~\cite{awesome_chatgpt_prompts, Prompt_Coder}. For each $p_{i}$, we generate $x_i$ by prompting LLMs with instructions aligned with $p_{i}$'s purpose. For example, if $p_{i}$ is ``\texttt{Generate a [product] advertisement\ldots}'', the LLM is prompted with ``\texttt{Provide an example of a product name}'' to generate $x_i = \texttt{smartphone}$. $y_i$ is then obtained by querying the LLMs using the combination of $x_i$ and $p_i$.

We then employ an LLM as the generative model $G$ to generate stolen prompt $p_{si}$ corresponding to $p_{i}$. We aim to analyze the key factors, such as theme, tone, and style, that influence the functional consistency between $p_{i}$ and $p_{si}$. Formally, this process is as follows:

\begin{equation}
a^{*} = \mathop{argmax}_{a}~\mathbb{E}_{(x_{i}, y_{i}) \in D_{c}} \left[ M(y_{i}, y_{si}) \right],
\label{eq:optimal_goal}
\end{equation}
\begin{align*}
        \text{s.t.} \ y_{i} = F_{t}(p_{i}, x_{i}),\ y_{si} = F_{t}(p_{si}, x_{i}), \ p_{si}=G(x_i, y_i, a),
\end{align*} 

\noindent where $a$ represents the above-mentioned key factors. The purpose of solving for $a$ is to ensure that $G$, when generating stolen prompts, focuses specifically on how the output content is expressed in $a$. So, we denote $a$ as \emph{prompt attention} for clarity. $\mathbb{E}_{(x_{i}, y_{i})}$ denotes the expected value across all pairs $(x_{i}, y_{i})$, and $y_{si}$ denotes the output generated by $p_{si}$. 

Equation~\ref{eq:optimal_goal} defines determining the attention as a single-objective optimization problem. Due to the complexity of the optimization landscape and the non-convex nature of LLMs, $a^{*}$ is denoted as an approximately optimal solution, typically addressed through heuristic or gradient-based methods. However, the target LLM $F_{t}$ is often black-box; even when open-source, gradient-based optimization is extremely challenging. Drawing inspiration from textual ``gradient'' in automatic prompt engineering~\cite{pryzant2023automatic, yang2023large} and the findings from Figure~\ref{fig:t_sne}, we propose a prompt attention algorithm based on the output differences. We outline the details of this algorithm in Algorithm~\ref{alg:differential_feedback_prompt_attention}.

\begin{algorithm}[t] \footnotesize
	\caption{Prompt attention algorithm based on the output differences.}
	\begin{algorithmic}[1]
		\Require{Input-output pairs $(x_{i}, y_{i})$ in $D_{c}$ with the same category, the generative model $G$, the target LLM $F_{t}$, and the number of iterations $N$.}
		\State Initialize attention $a = \{\}$ 
            \State Initialize stolen prompt $p_{si} = $`` ''
        \For {$n=0$ to $N-1$}
            \For {each $(x_{i}, y_{i})$ in $D_{c}$} 
                \If {$a == \{\}$}
                    \State $p_{si} = G(x_{i}, y_{i})$
                \Else
                    \State $p_{si} = G(x_{i}, y_{i}, a)$
                \EndIf
                \State $y_{si} = F_{t}(p_{si}, x_{i})$ 
                \State Compute output differences $d_{m}$ and impact scores $loss_{d_{m}}$: $\{d_{1}:loss_{d_{1}}, \ldots, d_{m}:loss_{d_{m}}\} = F_{t\nabla}(y_{si}, y_{i})$ 
                \State Retain $d_{m}$ where $loss_{d_{m}} > \theta_{a}$ and update $a = a \cup {d_{m}}$    
            \EndFor
        \EndFor \\
        \Return $a$ 
        
	\end{algorithmic}
	\label{alg:differential_feedback_prompt_attention}
\end{algorithm}

The description below covers a complete optimization iteration. Initially, the generative model $G$ generates an initial stolen prompt $p_{si}$ from the single input-output pair $(x_{i}, y_{i})$ (refer to line $6$). To evaluate $p_{si}$'s performance, we use both $p_{si}$ and $x_{i}$ as inputs for the target LLM $F_{t}$, which then produces output $y_{si}$ (refer to line $10$). Subsequently, a difference analysis is performed between $y_{si}$ and $y_i$, conducted by $F_{t}$. Specifically, we manually instruct $F_t$ to analyze the semantic, syntactic, and structural differences between $y_{si}$ and $y_i$, identifying the specific factors contributing to these differences (see Appendix~\ref{appendix:prompt_generation_instructions} for implementation details). We represent each factor as $d_{m}$. We then instruct $F_t$ to score $d_{m}$, quantifying their impact on the output differences as $loss_{d_{m}}$, similar to gradient loss in deep learning models (refer to line $11$).

Considering the variability of LLM outputs, $y_{si}$ in each iteration may cause shifts in different linguistic factors, making the resulting $d_m$ unstable. To mitigate this, our method performs multiple optimization iterations and retains only high-confidence $d_m$ values. Specifically, factors $d_{m}$ with $loss_{d_{m}}$ values exceeding a pre-set threshold $\theta_{a}$ are retained in $a$ for updates in subsequent iterations. As a result, with each iteration, $a$ accumulates factors $d_{m}$ that significantly contribute to the output differences. In the next iteration cycle, $a$ serves as feedback to guide $G$ in focusing on the expression of factor $d_{m}$ in the $(x_{i}, y_{i})$ to optimize the new $p_{ si}$ generation (refer to line $8$). After $N$ iterations, the algorithm generates an approximately optimal $a^*$ for the current category. Since the prompt generation phase is specific to the category of the target prompt, only the dataset from this category is required to generate $a^*$, allowing for offline execution.

The optimized $a^*$ is then fed back into $G$ to generate a $p_{s}$ that matches the functionality of the target prompt $p_{t}$ within the same category. The process of generating $p_{s}$ is expressed as follows:

\begin{equation}
        p_{s} = G(x_{t}, y_{t}, a^{*}),
\label{eq:define_ps}
\end{equation}

\noindent where $(x_{t}, y_{t})$ denotes the input-output data of $p_{t}$. Specifically, we first analyze $y_{t}$'s expression about each key factor $a_i$ in $a^*$ by asking $G$. Note that the query can be designed as ``\texttt{What is the {$a_{i}$} of the output in one sentence?}'' The responses are then used as \emph{instruction characteristics} of $p_{t}$ and fed back to $G$. $G$ subsequently generates $p_{s}$ with the same functionality as $p_{t}$. The instruction for generating $p_{s}$ could be designed as: ``\texttt{Your task is to generate an instruction based on the provided user input and output. The instruction should focus on the specified instruction characteristics.}'' Moreover, considering the interactivity of LLM applications, we can directly query descriptions of $p_{t}$ to articulate better the characteristics of $p_{t}$.

\subsection{Prompt Pruning}
In this phase, our goal is to filter out words $R$ in $p_{s}$ that are highly related to the user input to ensure that $p_{s}$ can achieve functional consistency with $p_t$ when dealing with various inputs. We define the formal optimization goal as follows:

\begin{equation}
        R^{*}=\mathop{argmax}\limits_{R}~M(F_{t}(mask(p_{s}, R), x_{ti}), y_{ti}),~~ \forall i \in \mathbb{Z}^+,
	\label{eq:multi-objective}
\end{equation}
\begin{align*}
        \text{s.t.} \ p_{s} = G(x_{t}, y_{t}, a),
\end{align*}

\noindent where $mask(\cdot)$ denotes an operation that uses placeholders ``$\{\}$'' to replace $R$ in $p_{s}$, the $(x_{ti}, y_{ti})$ denotes the different input-output pair for $p_{t}$.

To effectively filter out words in $p_{s}$ that are highly related to the input $x_{t}$, we propose a two-step strategy for automatically identifying related words. Initially, leveraging the insights from Section~\ref{sec:insight}, we employ a semantic similarity-based identification strategy to filter out the candidate set of words related to $x_t$ in $p_s$. We begin by identifying nouns in $x_t$ through part-of-speech tagging, forming a set $K$ as follows:

\begin{equation}
    K = \sum_{i=1}^{m} \left[ {w_{i} \mid POS(w_{i}) \in {``NOUN"}} \right],
    \label{eq:keywords-recognition}
\end{equation}

\noindent where $w_i$ denotes the words obtained after tokenizing $x_t$, and $POS$ is the part-of-speech tagging model~\cite{charniak1993equations}, sourced from Natural Language Toolkit (NLTK)~\cite{loper2002nltk}. Subsequently, we utilize the word2vec model pre-trained on the Google News dataset to compute the cosine similarity between words in $p_{s}$ and those in $K$~\cite{church2017word2vec}. We define the set $R$ of words related to the input based on this similarity, as described below:

\begin{equation}
    R =  \sum_{i=1}^{l} \sum_{j=1}^{n} \left[w_{p_j} \mid \frac{\vec{v}_{p_j} \cdot \vec{v}_{k_i}}{\|\vec{v}_{p_j}\| \|\vec{v}_{k_i}\|} \geq \gamma\right],
    \label{eq:related_words}
\end{equation}

\noindent where $w_{p_j}$ denote the words in $p_{s}$, $v_{p_j}$ and $v_{k_i}$ denote the vectors of words in $p_{s}$ and $K$ respectively, and $\gamma$ is the similarity threshold. $\gamma$ ensures that only words with a similarity score exceeding this threshold are included in the candidate set $R$, which is ranked by semantic similarity.

\begin{algorithm}[t] \footnotesize
	\caption{Selective beam search for related word identification.}
	\begin{algorithmic}[1]
		\Require{Related words list $R$, truncation factor $\alpha$, beam size $b$, evaluation frequency $e$, evaluation function $f_{n}$.}
		\State Initialize $beams = []$
		\State Iteration $S = \min(\alpha \times \text{len}(R), \text{len}(R))$
            \State Evaluation interval $s = \max(1, S/e)$
		\For {$i = 1$ to $S$ step $s$}
		    \State $current~beam = R[:i]$
		    \State $score = f_{n}(current~beam)$ \Comment{Function evaluates semantic similarity between outputs}
		    \State Append $(score, current~beam)$ to $beams$
		    \State Sort $beams$ in descending order by score
		    \State $beams = beams[:b]$ \Comment{Keep top $b$ beams}
		\EndFor \\
		\Return $beams[0][1]$ \Comment{Return the best beam}
	\end{algorithmic}
	\label{alg:selective_beam_search}
\end{algorithm}

Given that not all words in $R$ are relevant and excessive filtering out may reduce $p_s$'s functionality, we implement a selective beam search algorithm to refine $R$ and identify words highly related to $x_{t}$ (Algorithm~\ref{alg:selective_beam_search}). The process begins by initializing an empty list $beams$ to store potential word lists. To manage computational costs, the search space is constrained by a truncation factor $\alpha$ (refer to line $2$). Each search iteration involves selecting words sequentially from $R$ and assessing them via the evaluation function $f_{n}$. The evaluation involves masking the identified related words in $p_s$ and generating the output $y_s$ from this modified $p_s$. We then evaluate the semantic similarity between $y_s$ and the target output $y_t$ (refer to line $6$), forming the beam search score. Evaluations are limited by frequency $e$ to manage costs. The generated scores and corresponding word lists are added to $beams$, which are then sorted in descending order based on their scores. Only the top $b$ word lists, where $b$ is the beam size, are retained. The algorithm ultimately returns the phrase with the highest score from $beams$, indicating the word most highly related to $x_t$.

Based on this two-step strategy, we identify words in $p_s$ highly related to $x_t$ and mask them with placeholders. We note that some target prompts involve multiple user input fields with varying semantic relevance. As a result, the placeholders do not strictly follow a one-to-one mapping with these fields. We leverage the LLM's contextual understanding to associate placeholders with appropriate user inputs during inference, even when their numbers differ. This flexible strategy eliminates the need for predefined input schemas. This process enhances the generality of $p_s$, ultimately becoming the final stolen prompt generated by \system.

\section{Experiment Setup}
\label{sec:setup}

\subsection{Datasets}
\label{sec:Datasets}

\paragraphbe{Dataset for Prompt Generation Phase}
In the prompt generation phase, we collect a prompt dataset $D$ consisting of approximately $50$ prompts from each of the $18$ popular categories: Advertisements (Ads), Business, Code, Data, Email, Fashion, Food, Games, Health, Ideas, Language, Music, SEO, Sports, Study, Translation, Travel, and Writing, with each prompt accompanied by a corresponding input-output pair. These categories are selected based on popularity to ensure practical relevance. The dataset is sourced from open-source platforms~\cite{awesome_chatgpt_prompts, Prompt_Coder}.

During the evaluation tasks, we focus on assessing the effectiveness of attacks in two mainstream real-world prompt services: prompt marketplaces and LLM application stores.

\paragraphbe{Prompt Marketplaces} 
From PromptBase~\cite{PromptBase}, a leading prompt marketplace hosting over $130,000$ prompts, we randomly purchase $360$ prompts currently for sale to evaluate attack efficacy. We select these prompts from the $18$ popular categories as mentioned above. Each category includes $10$ prompts based on GPT-3.5 and $10$ on GPT-4, with each prompt showcasing one input-output pair example.

\paragraphbe{LLM Application Stores}
We conduct attack tests on $100$ popular GPTs from OpenAI's GPT Store~\cite{GPTStore}, chosen to align with our threat model, as they claim to include protective instructions against prompt leakage. To construct the input-output data, we pose questions to the GPTs likely to demonstrate their capabilities as input data. The resulting input-output pairs are collected through these interactions.

\subsection{Models}
\label{sec:Models}

\paragraphbe{Target LLM}
The target LLM is the model used in prompt services, which currently predominantly employs two versions: GPT-3.5 and GPT-4~\cite{OpenAI}. Our research accordingly focuses on both of these models as target LLMs. Specifically, for attacks on PromptBase, we use both GPT-3.5 and GPT-4 as the target LLMs. For attacks targeting GPTs, we specifically use GPT-4 as the target LLM.

\paragraphbe{Generative Model}
Generative models in our work are utilized to generate stolen prompts from the input-output data of target prompts. By default, the generative model is the same as the target LLM.

Detailed parameter settings of \system are provided in Appendix~\ref{sec:Parameter_Settings}.

\subsection{Baseline Methods}
\label{sec:Baseline Methods}

We adopt four different baselines:

\noindent \textbf{Automatic Prompt Engineering:} OPRO~\cite{yang2023large}. OPRO is a SOTA method for automatic prompt engineering. We use the original code with two minimal modifications (see Appendix~\ref{sec:details_of_opro} for details).

\noindent \textbf{Prompt Stealing Attack-1:} Sha et al.~\cite{sha2024prompt}. Since the code is not open source, we enumerate all prompt types based on the settings described in their paper. We then instruct the LLM to reconstruct the prompt according to the instructions in their paper and select the most effective one as the final reconstructed prompt.

\noindent \textbf{Prompt Stealing Attack-2:} output2prompt~\cite{zhang2024extracting}. We use the original code. Notably, for the attack evaluation on prompt marketplaces, we use the embedding of a given single output example as the model's input.

\noindent \textbf{Prompt Leaking Attack:} PLEAK~\cite{hui2024pleak}. PLEAK is a SOTA method for prompt leaking attacks. Considering the limitations of threat scenarios, we exclusively use the original code to evaluate its performance in LLM application stores in our experiments.

Note that for baseline methods involving LLMs (e.g., OPRO and Sha et al.), we ensure consistency using the same LLM version as \system. output2prompt trains the inversion model without direct LLM involvement. For PLEAK, which uses open-source shadow LLMs to generate adversarial queries, we use the same shadow LLMs specified in their paper to ensure fairness.

\subsection{Metrics}
\label{sec:Metrics}

\paragraphbe{Functional Consistency}
To evaluate the functional consistency between stolen prompts and target prompts, we assess their output similarity across three dimensions: semantic, syntactic, and structural.

We gauge semantic similarity using the cosine similarity between embedding vectors of the outputs after they are transformed using a sentence transformer~\cite{sentence_transformers}. The syntactic similarity is measured using FastKASSIM~\cite{chen2022fastkassim, boghrati2018conversation}, which pairs and averages the most similar parsing trees between documents. Structural similarity is assessed using the reciprocal of Jensen-Shannon (JS) divergence~\cite{thompson2015performance, Gretel}, with lower JS divergence indicating closer structural distribution. For consistency, we use the reciprocal of JS divergence as the measure of structural similarity.

We propose a score for the similarity between outputs based on semantic, syntactic, and structural similarity measures. Formally, its expression is as follows:

\begin{equation}
    Score = \frac{m-1}{2mn} \sum_{i=1}^{n} \sum_{j=1}^{m} \frac{ \sum_{k=1}^{m} M(y_{s_{ij}}, y_{ik}) }{ \sum_{k=j+1}^{m}  M(y_{ij}, y_{ik}) },
    \label{eq:Similarity_score}
\end{equation}

\noindent where $M$ is the evaluation metric—BLEU, FastKASSIM, or the reciprocal of JS divergence—yielding semantic, syntactic, or structural similarity scores, respectively. $y_{s_{ij}}$ is the output from the stolen prompt, while $y_{ij}$ and $y_{ik}$ are outputs from the target prompt. Due to the inherent randomness in LLM-generated content, we set a sampling number $m$ to repeatedly generate outputs for the same input, ensuring reliable evaluation. $n$ represents the number of test inputs to assess the stolen prompts' effectiveness. In our experiments, $m$ is set to $3$ and $n$ to $2$. We normalize our results against the similarity $M(y_{ij}, y_{ik})$ between target outputs, producing a similarity score $Score$ in the range $(0, 1)$, where scores closer to $1$ indicate higher similarity between stolen and target prompt outputs.

\paragraphbe{LLM-based Multi-dimensional Evaluation} 
We also conduct an LLM-based multidimensional evaluation of the attack performance. Specifically, we assess functional consistency across five comprehensive dimensions: \textit{Accuracy}, \textit{Completeness}, \textit{Tone}, \textit{Sentiment}, and \textit{Semantics}. For each output pair, GPT-4 assigns a score from $1$ to $10$ for each dimension, where higher values indicate better alignment.

\paragraphbe{Prompt Similarity} 
To further evaluate the functional consistency, we focus on analyzing the semantic similarity between the stolen and target prompts~\cite{hui2024pleak, sha2024prompt}, as this approach better captures whether the functional intent of the prompts remains consistent. We employ a sentence transformer to generate embeddings for these prompts and compute their similarities~\cite{sentence_transformers}.

\paragraphbe{Human Evaluation}
To validate the effectiveness of our stolen prompts from a human perspective, we conduct a human evaluation. We engage a diverse group of $20$ participants, including $10$ PhD holders in information security and $10$ general users, all of whom volunteer for the study. We randomly select $50$ target prompt cases from PromptBase~\cite{PromptBase} for evaluation. Participants respond to two questions: \textbf{Similarity:} Rate the similarity between outputs from the stolen and target prompts on a scale from $1$ to $10$, where a higher score indicates greater similarity. \textbf{Difference:} Identify and describe differences in the outputs based on their respective criteria.

Note that participants are blinded to the source of the outputs, and all outputs are anonymized. For each target prompt, the order of outputs from different methods is independently randomized to mitigate potential ordering bias.

\paragraphbe{Attack Success Rate}
We introduce the Attack Success Rate (ASR) to evaluate attack effectiveness. An attack is deemed successful if it meets all the following criteria: the semantic similarity score exceeds or equals threshold $\gamma_{sem}$, the syntactic similarity score exceeds or equals threshold $\gamma_{syn}$, and the structural similarity score exceeds or equals threshold $\gamma_{str}$.

\section{Attack Performance on Prompt Marketplaces}
\label{sec:Performance_in_Prompt_Marketplaces}

In this section, we thoroughly evaluate the performance of \system and the baseline methods in conducting attacks on real-world prompt marketplaces.

\paragraphbe{Similarity Analysis of Prompts $p_{i}$ in $D$ and Target Prompts $p_{t}$}
Before evaluating the attack performance, we assess the similarity between prompts $p_{i}$ in $D$ and target prompts $p_{t}$. Our results show that the data in $D$ is not highly similar to the target prompts. In addition, we further evaluate \system's performance under extreme conditions where prompts in $D$ are either highly similar or deliberately dissimilar to the target prompts. Our findings suggest that \system’s effectiveness is not strongly dependent on the similarity between the auxiliary data and the target prompts. See Appendix~\ref{appendix:similarity_analysis_with_Dc} for details.

\subsection{Functional Consistency}
\label{sec:Effectiveness_task1}

\begin{table*}
    \centering
    \caption{Attack performance of \system and baseline methods in stealing GPT-3.5 based prompts from prompt marketplaces. The best results are shown in \textbf{bold}. ``--'' indicates not applicable.}
    \label{tab:performance_on_gpt3_based_prompt}
    \begin{adjustbox}{width=\textwidth}
    \renewcommand{\arraystretch}{1.1}
    \setlength{\tabcolsep}{0.6mm}
    \begin{tabular}{cccccccccccccccccccc}
        \toprule
        \multirow{3}{*}{\textbf{\normalsize Metric}}         & \multirow{3}{*}{\textbf{\normalsize Attack Method}} & \multicolumn{18}{c}{\textbf{Category}}                               \\ \cmidrule(l){3-20} 
                                &                         & \textbf{Ads} & \textbf{Business} & \textbf{Code} & \textbf{Data} & \textbf{Email} & \textbf{Fashion} & \textbf{Food} & \textbf{Games} & \textbf{Health}
                                & \textbf{Ideas}  & \textbf{Language} & \textbf{Music}& \textbf{SEO}& \textbf{Sports}
                                & \textbf{Study} & \textbf{Translation} & \textbf{Travel} & \textbf{Writing} \\ \midrule
        \multirow{6}{*}{\begin{tabular}[c]{@{}c@{}}{Semantic} \\ {Similarity}\end{tabular}} 
        & OPRO & 0.49 & 0.53 & 0.51 & 0.59 & 0.59 & 0.50 & 0.61 & 0.62 & 0.50 & 0.62 & 0.48 & 0.63 & 0.42 & 0.63 & 0.49 & 0.28 & 0.51 & 0.55 \\
        & Sha et al. & 0.49 & 0.50 & 0.45 & 0.61 & 0.43 & 0.62 & 0.57 & 0.64 & 0.60 & 0.60 & 0.53 & 0.63 & 0.50 & 0.69 & 0.54& 0.46 & 0.60 & 0.56 \\
        & output2prompt & 0.52 & 0.53 & 0.56 & 0.63 & 0.50 & 0.61 & 0.62 & 0.62 & 0.56 & 0.48 & 0.43 & 0.59 & 0.55 & 0.55 & 0.58 & 0.28 & 0.61 & 0.56 \\
        & PLEAK & -- & -- & -- & -- & -- & -- & -- & -- & -- & -- & -- & -- & -- & -- & -- & -- & -- & --\\
         & \system & \textbf{0.70} & \textbf{0.73} & \textbf{0.61} & \textbf{0.80} & \textbf{0.75} & \textbf{0.83} & \textbf{0.73} & \textbf{0.83} & \textbf{0.75} & \textbf{0.85} & \textbf{0.70} & \textbf{0.86} & \textbf{0.75} & \textbf{0.83} & \textbf{0.67} & \textbf{0.74} & \textbf{0.79} & \textbf{0.71}\\
        \cmidrule(lr){2-20}
        & \% Gain for \system & 34.62 & 37.74 & 8.93 & 26.98 & 27.12 & 33.87 & 17.74 & 29.69 & 25.00 & 37.10 & 32.08 & 36.51 & 36.36 & 20.29 & 15.52 & 60.87 & 29.51 & 26.79 \\
        \midrule
        
        \multirow{6}{*}{\begin{tabular}[c]{@{}c@{}}{Syntactic} \\ {Similarity}\end{tabular}} 
        & OPRO & 0.66 & 0.59 & 0.53 & 0.57 & 0.53 & 0.42 & 0.52 & 0.64 & 0.42 & 0.28 & 0.57 & 0.65 & 0.51 & 0.63 & 0.80 & 0.31 & 0.75 & 0.65\\
        & Sha et al. & 0.57 & 0.50 & 0.41 & 0.62 & 0.52 & 0.68 & 0.70 & 0.74 & 0.62 & 0.53 & 0.41 & 0.78 & 0.56 & 0.65 & 0.72 & 0.33 & 0.76 & 0.59 \\
        & output2prompt & 0.68 & 0.34 & 0.65 & 0.45 & 0.32 & 0.58 & 0.56 & 0.48 & 0.49 & 0.35 & 0.39 & 0.21 & 0.47 & 0.29 & 0.68 & 0.15 & 0.56 & 0.47 \\
        & PLEAK & -- & -- & -- & -- & -- & -- & -- & -- & -- & -- & -- & -- & -- & -- & -- & -- & -- & -- \\
        & \system & \textbf{0.91} & \textbf{0.79} & \textbf{0.75}  & \textbf{0.83} & \textbf{0.90} & \textbf{0.89} & \textbf{0.86} & \textbf{0.88} & \textbf{0.86} & \textbf{0.79} & \textbf{0.76} & \textbf{0.91} & \textbf{0.61} & \textbf{0.89} & \textbf{0.91} & \textbf{0.73} & \textbf{0.89} & \textbf{0.74}\\
        \cmidrule(lr){2-20}
        & \% Gain for \system & 33.82 & 33.90 & 15.38 & 33.87 & 69.81 & 30.88 & 22.86 & 18.92 & 38.71 & 49.06 & 33.33 & 16.67 & 8.93 & 36.92 & 13.75 & 121.21 & 17.11 & 13.85\\
        \midrule

        \multirow{6}{*}{\begin{tabular}[c]{@{}c@{}}{Structural} \\ {Similarity}\end{tabular}} 
        & OPRO & 0.85 & 0.81 & 0.50 & 0.59 & 0.79 & 0.69 & 0.76 & 0.76 & 0.73 & 0.81 & 0.80 & 0.82 & 0.72 & 0.75 & 0.81 & 0.35 & 0.85 & 0.79 \\
        & Sha et al. & 0.81 & 0.72 & 0.59 & 0.84 & 0.75 & 0.79 & 0.81 & 0.81 & 0.78 & 0.81 & 0.75 & 0.85 & 0.74 & 0.82 & 0.84 & 0.54 & 0.85 & 0.76 \\
        & output2prompt & 0.76 & 0.63 & 0.71 & 0.71 & 0.67 & 0.81 & 0.77 & 0.80 & 0.77 & 0.58 & 0.79 & 0.69 & 0.71 & 0.73 & 0.83 & 0.21 & 0.82 & 0.76 \\
        & PLEAK & -- & -- & -- & -- & -- & -- & -- & -- & -- & -- & -- & -- & -- & -- & -- & -- & -- & -- \\
        & \system & \textbf{0.89} & \textbf{0.85} & \textbf{0.87} & \textbf{0.91} & \textbf{0.91} & \textbf{0.86} & \textbf{0.91} & \textbf{0.95} & \textbf{0.87} & \textbf{0.89} & \textbf{0.87} & \textbf{0.92} & \textbf{0.80} & \textbf{0.93} & \textbf{0.94} & \textbf{0.75} & \textbf{0.92} & \textbf{0.83} \\
        \cmidrule(lr){2-20}
        & \% Gain for \system & 4.71 & 4.94 & 22.54 & 8.33 & 15.19 & 6.17 & 12.35 & 17.28 & 11.54 & 9.88 & 8.75 & 8.24 & 8.11 & 13.41 & 11.90 & 38.89  & 8.24 & 5.06 \\
        \bottomrule
    \end{tabular}
    \end{adjustbox}
\end{table*}

We evaluate the functional consistency of \system and baseline methods in stealing GPT-3.5 based prompts. The results in Table~\ref{tab:performance_on_gpt3_based_prompt} show that \system outperforms the baselines. Notably, PLEAK, which relies on interaction with target LLMs, is not applicable in prompt marketplaces.

We analyze semantic similarity, where \system achieves the highest average score across all categories at $0.76$, compared to $0.56$ for the best baseline. Regarding syntactic similarity, \system also leads, with over a $30.0\%$ gain in $10$ out of $18$ categories, surpassing the baseline by more than $50.0\%$ in the ``Email'' and ``Translation'' categories. For structural similarity, \system shows a $14.3\%$ average improvement. Further evaluations on GPT-4 based prompts show similar trends, detailed in Appendix~\ref{appendix:Effectiveness_gpt_4}. We also include a sample size sensitivity analysis in Appendix~\ref{appendix:sensitivity_analysis}. Overall, \system demonstrates notable improvements   over baselines.

\begin{table}[h]
    \footnotesize
    \centering
    \caption{Prompt similarity between target and stolen prompts.}
    \label{tab:prompt_similarity}
    \setlength{\tabcolsep}{0.6mm}
    \begin{tabular}{cccccc}
        \toprule
        \multirow{3}{*}{\textbf{Metric}}         & \multirow{3}{*}{\textbf{Target Prompt}} & \multicolumn{4}{c}{\textbf{Attack Method}}                               \\ \cmidrule(l){3-6} 
                                &                         & OPRO & Sha et al. & output2prompt & \system \\ \midrule
        \multirow{2}{*}{\begin{tabular}[c]{@{}c@{}}{\scriptsize{Prompt}} \\ {\scriptsize{Similarity}}\end{tabular}} 
        & \scriptsize{GPT-3.5 Based Prompt} & 0.45 & 0.45 & 0.34 & \textbf{0.69}\\
        & \scriptsize{GPT-4 Based Prompt} & 0.50 & 0.52 & 0.34 & \textbf{0.73}\\
        
        \bottomrule
    \end{tabular}
\end{table}

To further evaluate the functional consistency, we also compare the prompt similarity between stolen and target prompts. Table~\ref{tab:prompt_similarity} shows that \system generates stolen prompts more similar to the targets than baseline methods, with over a $53.3\%$ improvement when targeting GPT-3.5 based prompts. Furthermore, methods that rely on LLM capabilities, such as OPRO, Sha et al., and \system perform better on GPT-4 based prompts than on GPT-3.5, indicating that advancements in generative models enhance stealing attacks.

\subsection{LLM-based Multi-dimensional Evaluation}

\begin{table}
    \footnotesize
    \centering
    \caption{LLM-based multi-dimensional evaluation of functional consistency. Higher scores indicate better alignment.}
    \label{tab:llm_based_evaluation}
    \setlength{\tabcolsep}{0.8mm}
    \begin{tabular}{cccccc}
        \toprule
        \multirow{2}{*}{\textbf{Target Prompt}} & \multirow{2}{*}{\textbf{Metric}} & \multicolumn{4}{c}{\textbf{Attack Method}} \\ \cmidrule(l){3-6} 
         &  & OPRO & Sha et al. & output2prompt & \system \\ \midrule
        \multirow{4}{*}{\scriptsize{\makecell[c]{GPT-3.5 \\ Based Prompt}}} 
         
         & \scriptsize{Accuracy}     & 3.62 & 3.64 & 4.73 & \textbf{7.04} \\
         & \scriptsize{Completeness} & 3.28 & 3.31 & 4.32 & \textbf{7.10} \\
         & \scriptsize{Semantics}    & 4.25 & 3.76 & 4.83 & \textbf{7.63} \\
         & \scriptsize{Sentiment}    & 7.61 & 7.34 & 7.59 & \textbf{9.15} \\
         & \scriptsize{Tone}         & 7.59 & 6.94 & 7.14 & \textbf{9.18} \\ \midrule
        \multirow{4}{*}{\scriptsize{\makecell[c]{GPT-4 \\ Based Prompt}}} 
         & \scriptsize{Accuracy}     & 5.56 & 5.86 & 5.14 & \textbf{7.36} \\
         & \scriptsize{Completeness} & 5.74 & 5.83 & 4.92 & \textbf{7.58} \\
         & \scriptsize{Semantics}    & 6.17 & 6.16 & 5.62 & \textbf{8.06} \\
         & \scriptsize{Sentiment}    & 8.77 & 8.85 & 8.18 & \textbf{9.27} \\
         & \scriptsize{Tone}         & 8.86 & 8.84 & 8.14 & \textbf{9.32} \\ 
        \bottomrule
    \end{tabular}
\end{table}

Table~\ref{tab:llm_based_evaluation} shows the LLM-based evaluation across five dimensions. \system performs well across all dimensions, with noticeable improvements over the baselines in \textit{Accuracy}, \textit{Completeness}, and \textit{Semantics}. All baselines achieve higher scores on GPT-4 based prompts than on GPT-3.5 based prompts, which may be attributed to GPT-4’s stronger reasoning capabilities. Although output2prompt is model-agnostic, it still shows slight improvements on GPT-4 based prompts, possibly due to clearer output intent. These trends align with our functional consistency evaluation results.

\subsection{Human Evaluation}

\begin{figure}[h]
	\setlength{\abovecaptionskip}{1pt}
	\captionsetup[subfigure]{justification=centering}
	\centering
        \begin{subfigure}{0.475\linewidth}
		\includegraphics[width=\textwidth]{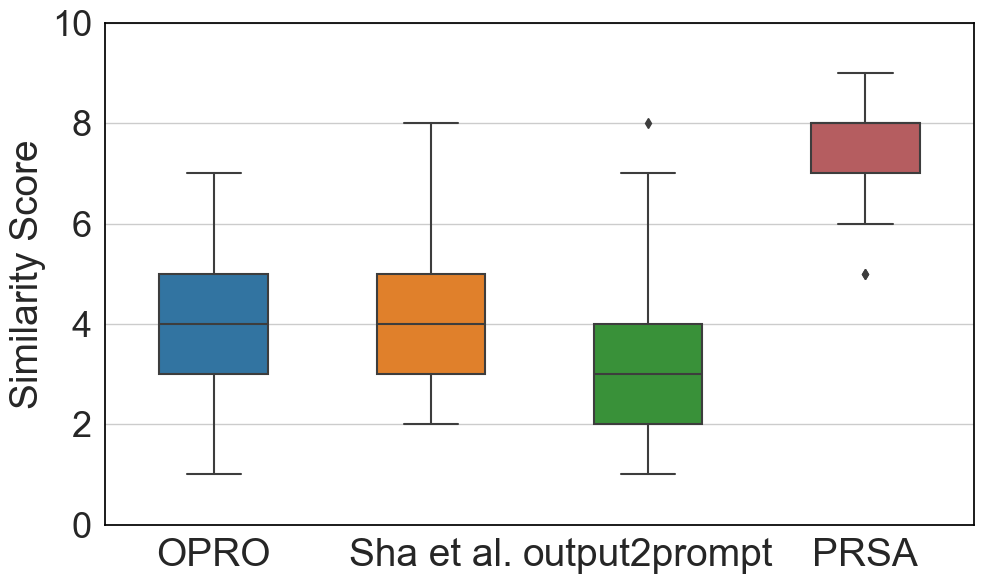}
		\caption{\small{GPT-3.5 Based Prompt}}\label{fig:huaman_evaluation_a}
	\end{subfigure}
	\hspace{0.2cm}
	\begin{subfigure}{0.475\linewidth}
		\includegraphics[width=\textwidth]{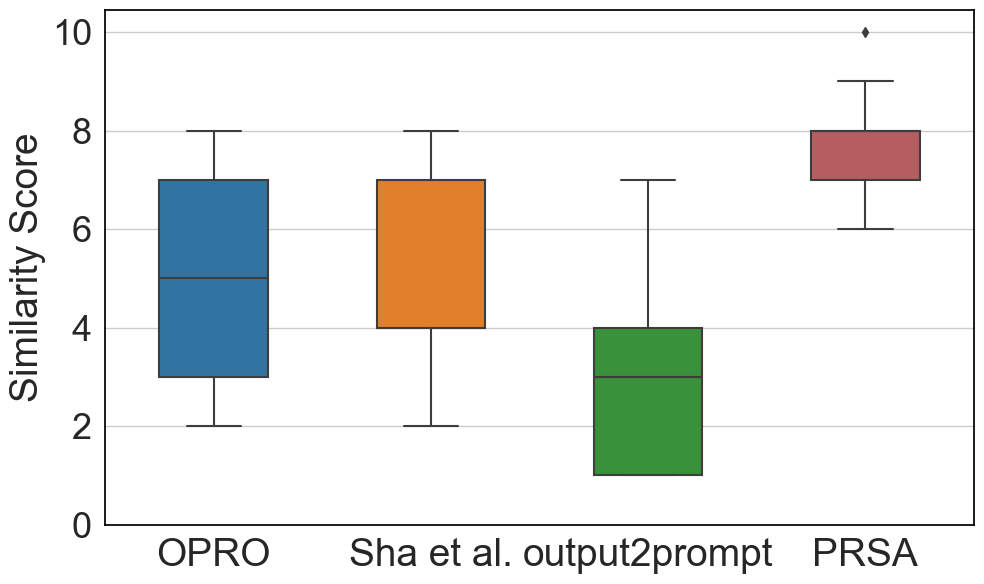}
		\caption{\small{GPT-4 Based Prompt}}\label{fig:huaman_evaluation_b}
	\end{subfigure}
        
  \caption{Similarity scores assessed by humans for outputs from target and stolen prompts. The stolen prompts are generated by \system and baseline methods.}
  \label{fig:huaman_evaluation}
\end{figure}

We measure the effectiveness of \system and baseline methods from a human perspective (see Section~\ref{sec:Metrics} for the evaluation setup). The results are illustrated in Figure~\ref{fig:huaman_evaluation}. Figure~\ref{fig:huaman_evaluation_a} displays the performance of stealing attacks based on GPT-3.5 prompts. The narrower Interquartile Range (IQR) for \system indicates a higher level of score consistency. Moreover, the median score of the \system is higher than the baseline methods. A similar trend is observed when comparing outputs from steal attacks based on GPT-4 prompts, as shown in Figure~\ref{fig:huaman_evaluation_b}.

Incorporating user feedback, we identify that semantic inconsistencies highlighted by most users are the primary reason for the lower performance of baseline methods. Another critical issue is that the stolen prompts generated by these baseline methods often include user input-related content, leading to inaccurate or irrelevant responses when applied to new input scenarios. Despite \system achieving the highest ratings, some users still point out its shortcomings, primarily regarding its ability to match the semantics of the target output.

\subsection{Attack Effectiveness}

To further quantify attack effectiveness, we explore the optimal functional consistency thresholds, aiming to meet the successful attack criteria defined in Section~\ref{sec:Metrics}. Experimental results reveal that the highest consistency between the defined criteria and actual outcomes is achieved with $\gamma_{sem}=0.75$, $\gamma_{syn}=0.75$, and $\gamma_{str}=0.9$. Further details can be found in Appendix~\ref{appendix:Threshold_Exploration_for_ASR}. The ASRs of different methods are shown in Table~\ref{tab:ASR_Non_interactive_service}. \system's ASR outperforms baselines, achieving $280\%$ higher ASR than baselines for GPT-3.5 based prompt.

\begin{table}[h]
    \footnotesize
    \centering
    \caption{Comparative ASR of \system and baseline methods.}
    \label{tab:ASR_Non_interactive_service}
    \setlength{\tabcolsep}{0.6mm}
    \begin{tabular}{cccccc}
        \toprule
        \multirow{3}{*}{\textbf{Metric}}         & \multirow{3}{*}{\textbf{Target Prompt}} & \multicolumn{4}{c}{\textbf{Attack Method}}                               \\ \cmidrule(l){3-6} 
                                &                         & OPRO & Sha et al. & output2prompt & \system \\ \midrule
        \multirow{2}{*}{ASR} 
        & \scriptsize{GPT-3.5 Based Prompt} & $4.4\%$      & $6.7\%$      & $8.3\%$     & $\textbf{31.7\%}$\\
        & \scriptsize{GPT-4 Based Prompt} & $17.2\%$      & $17.8\%$      & $8.9\%$     & $\textbf{46.1\%}$\\
        
        \bottomrule
    \end{tabular}
\end{table}

\paragraphbe{Discussion on Why Baselines Fail in Attacks Against Prompt Marketplaces} 
We analyze the reasons for the poor performance of each baseline by examining their effectiveness results and inherent characteristics. OPRO typically needs multiple input-output pairs to optimize prompts, but real-world prompt marketplaces usually provide only a single pair, limiting its effectiveness. Sha et al.'s method, which relies entirely on the backward inference capabilities of LLMs, struggles with the complex and rich semantics of commercial prompts. Additionally, both Sha et al.'s work and OPRO overlook the generality of stolen prompts. output2prompt aims to learn tailored input-output features to generate stolen prompts. Still, it falters as prompt marketplaces typically do not supply the necessary examples required by adversaries.

\paragraphbe{Cost Analysis}
The effectiveness of an attack depends on both its ASR and economic feasibility. To be profitable, the attack cost, including computing resources and querying LLM APIs, must be much lower than the cost of purchasing target prompts. Our results show that \system is highly cost-effective. Specifically, the average attack cost, calculated as the attack cost divided by the number of successfully attacked prompts, is $1.3\%$--$2.1\%$ of the average price for GPT-3.5 based prompts, and $11.6\%$--$12.3\%$ for GPT-4 based prompts. See Appendix~\ref{appendix:Cost_Analysis} for details.

\subsection{Ablation Study}
\label{sec:ablation_study_section}

\begin{table}
    \centering
    \footnotesize
    \setlength{\abovecaptionskip}{2pt}
    \caption{Evaluating the average effectiveness of \system through ablation study. PG denotes the prompt generation phase, while PP signifies the prompt pruning phase.}
    \setlength{\tabcolsep}{3.1mm}
    \begin{tabular}{ccccc}
        \toprule
        \multirow{3}{*}{\textbf{Attack Method}} & \multicolumn{3}{c}{\textbf{Metric}} \\ 
        \cmidrule(l){2-4} 
        & \begin{tabular}[c]{@{}c@{}}{Semantic} \\{Similarity} \end{tabular} &
        \begin{tabular}[c]{@{}c@{}}{Syntactic} \\{Similarity} \end{tabular} &
        \begin{tabular}[c]{@{}c@{}}{Structural} \\{Similarity} \end{tabular} \\ 
        \midrule
        {\system w/o PG and PP} & 0.59 & 0.63 & 0.82 \\ 
        {\system w/o PG} & 0.71 & 0.78 & 0.87 \\
        {\system w/o PP} & 0.68 & 0.85 & 0.90 \\
        {\system} & \textbf{0.80} & \textbf{0.88} & \textbf{0.91} \\ 
        \bottomrule
    \end{tabular}
    \label{tab:ablation_study}
\end{table}

We conduct an ablation study using prompts from six common categories: ``Ads'', ``Email'', ``Ideas'', ``Music'', ``Sports'', and ``Travel''. As shown in Table~\ref{tab:ablation_study}, removing the prompt generation (PG) phase from the \system (retaining only the basic capabilities to generate stolen prompts) reduces semantic, syntactic, and structural similarity scores, highlighting its role in accurately inferring the target prompt's intent. Removing the prompt pruning (PP) phase leads to a $17.6\%$ drop in semantic similarity, demonstrating its importance in enhancing prompt generality. Thus, both PG and PP are essential for improving semantic similarity, with PG having a stronger impact on syntactic and structural aspects.

\section{Attack Performance on LLM Application Stores}
\label{sec:Performance_on_LLM_Applications}

\subsection{Attack Effectiveness}
\label{sec:Effectiveness_task2}

\begin{figure*}[t]
	\setlength{\abovecaptionskip}{1pt}
	\captionsetup[subfigure]{justification=centering}
	\centering
        \begin{subfigure}{0.31\linewidth}
		\includegraphics[width=\textwidth]{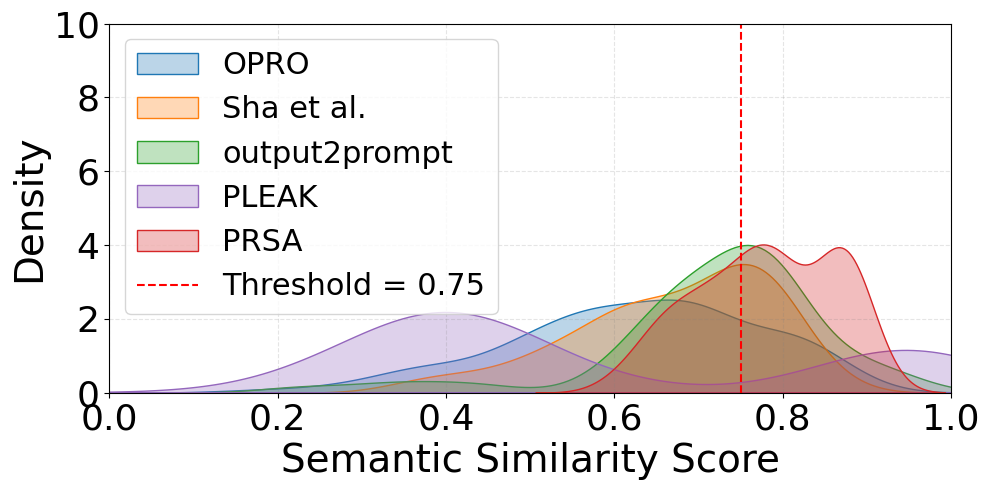}
		\caption{\small{Semantic Similarity}}\label{fig:ASR_query_size}
            \label{fig:gpts_sem}
	\end{subfigure}
	\hspace{0.2cm}
	\begin{subfigure}{0.31\linewidth}
		\includegraphics[width=\textwidth]{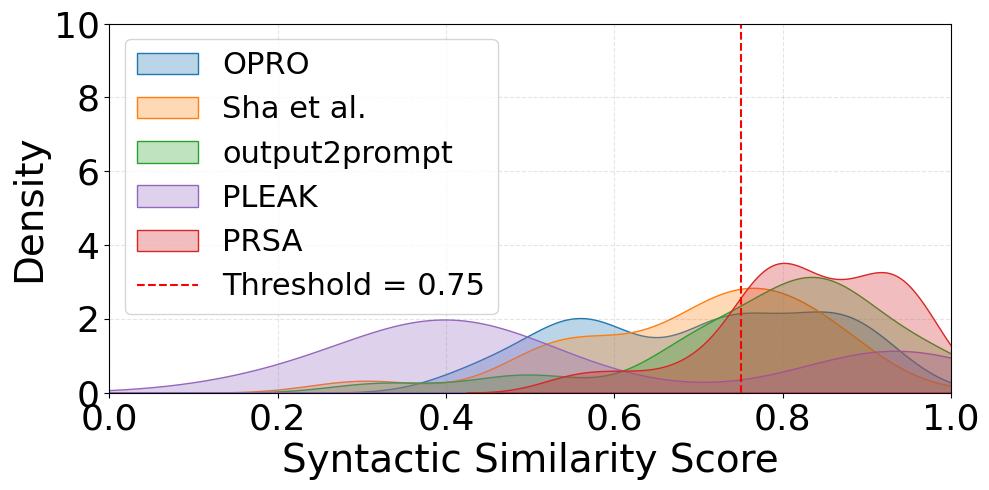}
		\caption{\small{Syntactic Similarity}}\label{fig:semantic_query_size}
            \label{fig:gpts_syn}
	\end{subfigure}
	\hspace{0.2cm}
	\begin{subfigure}{0.31\linewidth}
		\includegraphics[width=\textwidth]{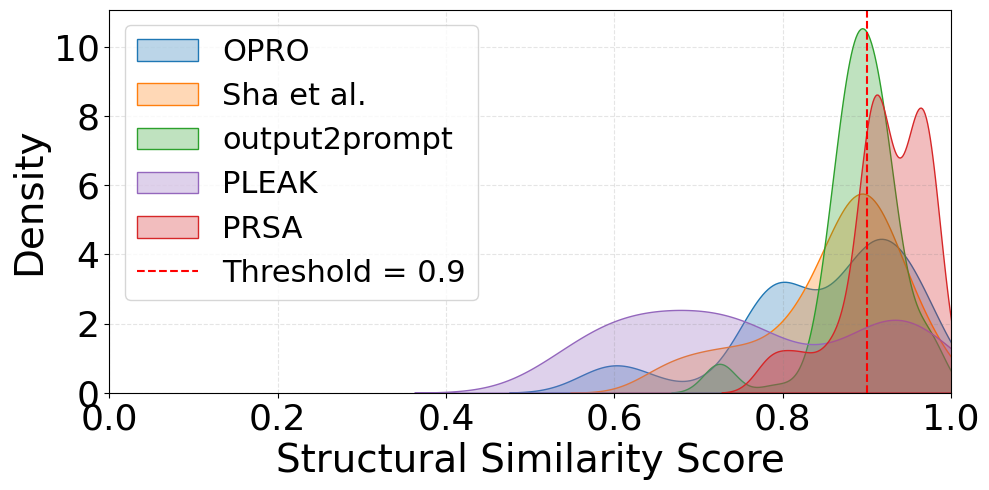}
		\caption{\small{Structural Similarity}}\label{fig:syntactic_query_size}
            \label{fig:gpts_str}
	\end{subfigure}
  \caption{Distribution of attack performance of \system and baseline methods in stealing system prompts from $100$ GPTs.}
  \label{fig:stealing_attack_on_gpts}
\end{figure*}
In this section, we evaluate the effectiveness of \system and baseline methods in stealing system prompts from GPTs equipped with protective instructions in OpenAI's official GPT Store. Figure~\ref{fig:stealing_attack_on_gpts} displays the attack performance distribution for \system and four baselines using kernel density estimates. \system achieves a high-density peak around a semantic similarity score of $0.8$ (Figure~\ref{fig:gpts_sem}), outperforming most baselines. Syntactic similarity (Figure~\ref{fig:gpts_syn}) shows that over $70\%$ of \system's scores exceed $0.75$. For structural similarity (Figure~\ref{fig:gpts_str}), \system peaks near the $0.9$ threshold. Table~\ref{tab:ASR_Interactive_service} shows that \system leads with an ASR of $52\%$, well above the best-performing baseline methods.

To understand why system prompts can still be stolen despite protections, we contact the developers of the evaluated GPTs. They confirm the presence of such protections, though their mechanisms remain undisclosed. Based on public examples~\cite{gpt_protections}, we speculate these protections block direct leakage but still allow outputs revealing prompt intent.

\begin{table}
    \footnotesize
    \centering
    \caption{Comparative ASR of \system and baseline methods for stealing system prompts of GPTs.}
    \label{tab:ASR_Interactive_service}
    \setlength{\tabcolsep}{2mm}
    \begin{tabular}{cccccc}
        \toprule
        \multirow{3}{*}{\textbf{Metric}}     & \multicolumn{5}{c}{\textbf{Attack Method}}                               \\ \cmidrule(l){2-6} 
                                & OPRO & Sha et al. & output2prompt & PLEAK & \system \\ \midrule
        ASR
        
        & $16\%$      & $14\%$      & $39\%$  & $31\%$   & $\textbf{52\%}$\\
        
        \bottomrule
    \end{tabular}
\end{table}

\paragraphbe{Discussion on Why Baselines Fail in Attacks Against LLM Application Stores} 
We analyze the reasons for the poor performance of each baseline method in the context of our experiments. As LLM applications, GPTs feature system prompts with high generality, a challenge that OPRO and Sha et al. fail to address. output2prompt struggles with data drift~\cite{mallick2022matchmaker, ackerman2021automatically} as real-world system prompts often fall outside its training data distribution. Protective instructions direct LLMs to block queries related to system prompts, significantly reducing the effectiveness of prompt leaking attacks such as PLEAK. The bimodal distribution in Figure~\ref{fig:stealing_attack_on_gpts} captures this variability: PLEAK either fails to extract the system prompts or achieves outstanding success.

\subsection{Case Study}
\label{sec:Case_Study}

To clearly illustrate the utility of \system, we select two popular GPTs from OpenAI's GPT Store as case studies. Both GPTs claim to have protective instructions to prevent the leakage of system prompts:

\textbf{Global Travel Planner} \footnote{\scriptsize\url{https://chatgpt.com/g/g-veMpTb39A-global-travel-planner}}. Global Travel Planner is a travel planning application with over $10K$ conversations.

\textbf{Math} \footnote{\scriptsize\url{https://chatgpt.com/g/g-WP1diWHRl-math}}. Math is a ChatGPT-based math tool, which was previously ranked 3rd globally in the Education category and now has over $1M$ conversations.

For Global Travel Planner, Table~\ref{tab:Case_study_Global_Travel_Planner} shows that only \system meets the criteria for a successful attack. OPRO and Sha et al. perform poorly. We observe that Global Travel Planner typically outputs in Chinese, whereas output2prompt, trained on English data, fails to extract the system prompt and sometimes produces gibberish. Attempts to inject optimized adversarial instructions via PLEAK are unsuccessful. The stolen prompt generated by \system was confirmed by the developers of Global Travel Planner, further validating its effectiveness.

\begin{table}[h]
    \scriptsize
    \centering
    \caption{Performance evaluation of attacks on ``Global Travel Planner'' GPTs. \emph{Sem. Sim.} denotes semantic similarity, \emph{Syn. Sim.} denotes syntactic similarity, \emph{Str. Sim.} denotes structural similarity, and \emph{Att. Succ.} denotes attack success.}
    \label{tab:Case_study_Global_Travel_Planner}
    \setlength{\tabcolsep}{0.75mm}
    \begin{tabular}{cccccc}
        \toprule
        \multirow{3}{*}{\textbf{Metric}} & \multicolumn{5}{c}{\textbf{Attack Method}} \\ \cmidrule(l){2-6} 
                                & OPRO & Sha et al. & output2prompt & PLEAK & \system \\ \midrule
        \scriptsize{Sem. Sim.}   & 0.21 ($\pm$0.03)  & 0.25 ($\pm$0.03)  & 0.56 ($\pm$0.03)  & 0.55 ($\pm$0.03) & \textbf{0.83} ($\pm$0.03) \\
        \scriptsize{Syn. Sim.}  & 0.44 ($\pm$0.04) & 0.50 ($\pm$0.03)  & 0.20 ($\pm$0.03) & 0.16 ($\pm$0.03) & \textbf{0.77} ($\pm$0.03) \\
        \scriptsize{Str. Sim.} & 0.70 ($\pm$0.01) & 0.72 ($\pm$0.01)  & 0.77 ($\pm$0.01) & 0.77 ($\pm$0.01) & \textbf{0.91} ($\pm$0.01) \\
        \scriptsize{Att. Succ.}        & \ding{55} & \ding{55} & \ding{55} & \ding{55} & \ding{51} \\
        \bottomrule
    \end{tabular}
\end{table}

\begin{table}[h]
    \scriptsize
    \centering
    \caption{Performance evaluation of attacks on ``Math'' GPTs.}
    \label{tab:Case_study_Math}
    \setlength{\tabcolsep}{0.75mm}
    \begin{tabular}{cccccc}
        \toprule
        \multirow{3}{*}{\textbf{Metric}} & \multicolumn{5}{c}{\textbf{Attack Method}} \\ \cmidrule(l){2-6} 
                                & OPRO & Sha et al. & output2prompt & PLEAK & \system \\ \midrule
        \scriptsize{Sem. Sim.}   & 0.60 ($\pm$0.05)  & 0.57 ($\pm$0.05) & 0.70 ($\pm$0.03)  & 0.68 ({\color{red}$\pm$0.20}) & \textbf{0.85} ($\pm$0.05) \\
        \scriptsize{Syn. Sim.}  & 0.63 ($\pm$0.05) & 0.52 ($\pm$0.05) & 0.71 ($\pm$0.03) &  0.63 ({\color{red}$\pm$0.30}) & \textbf{0.89} ($\pm$0.05) \\
        \scriptsize{Str. Sim.} & 0.81 ($\pm$0.02) & 0.85 ($\pm$0.02)  & 0.93 ($\pm$0.02) & 0.81 ({\color{red}$\pm$0.14}) & \textbf{0.93} ($\pm$0.02) \\
        \scriptsize{Att. Succ.}        & \ding{55} & \ding{55} & \ding{55} & \ding{51} & \ding{51} \\
        \bottomrule
    \end{tabular}
\end{table}

For Math, Table~\ref{tab:Case_study_Math} shows that baselines such as OPRO and Sha et al. underperform. output2prompt achieves moderate scores, but it does not meet the threshold for a successful attack. The stolen prompt generated by output2prompt fails to capture the broader functionality of Math. Although PLEAK is successful, it shows instability due to protective instructions. In contrast, \system exhibits higher stability and has successfully carried out attacks. Interestingly, while inferring the target system prompt using \system, we discover an Easter egg hidden within Math: when in calculator mode, entering the number $4$ triggers an unfolding ASCII art narrative. This characteristic is not mentioned in Math's official introduction. Notably, OPRO, Sha et al., and output2prompt fail to detect this intriguing characteristic. The stolen prompt we obtained is also confirmed by Math's developers.

We create replicas of the above GPTs using the stolen prompts generated by \system (with permission from the developers) to demonstrate the threat of \system. These demo versions are anonymously available at\url{https://sites.google.com/view/prsa-prompt-stealing-attack}.

\section{Why Our Attacks Work}

To understand the prompt stealing attacks, we explore why \system can infer the target prompt using an input-output pair, particularly focusing on the output. Intuitively, if the output contains a lot of information about the target prompt, a prompt stealing attack is feasible. To quantify this, we introduce the concept of \emph{mutual information}, which quantifies the extent of information shared between the outputs and the target prompts. Theoretical analysis is provided in Appendix~\ref{appendix:Theoretical_Analysis}.

\begin{figure}
	\setlength{\abovecaptionskip}{1pt}
	\captionsetup[subfigure]{justification=centering}
	\centering
        \begin{subfigure}{0.475\linewidth}
		\includegraphics[width=\textwidth]{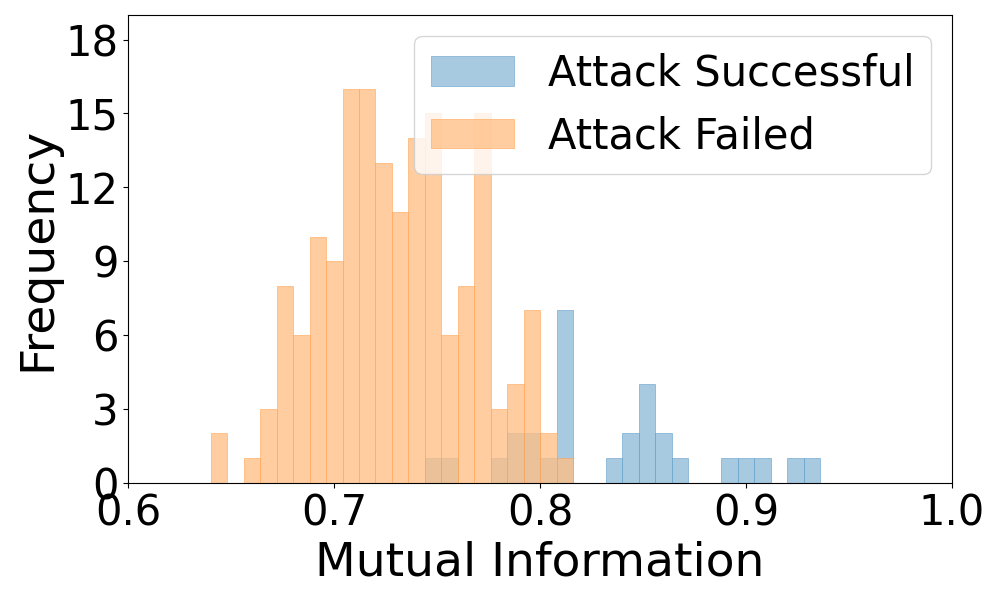}
		\caption{\small{\system w/o Prompt Attention}}\label{fig:PRSA_wo_attention}
	\end{subfigure}
	\hspace{0.2cm}
	\begin{subfigure}{0.475\linewidth}
		\includegraphics[width=\textwidth]{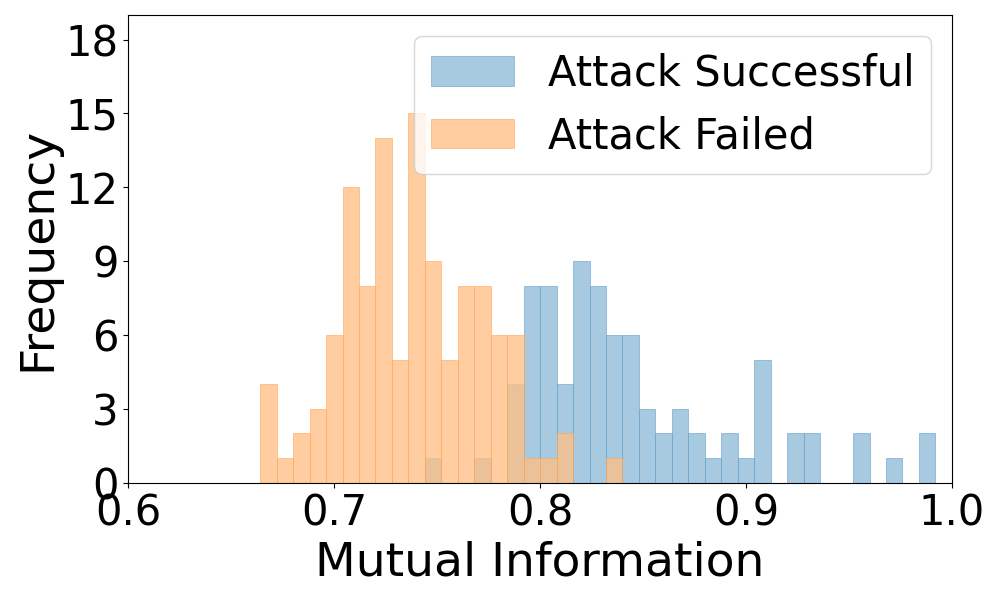}
		\caption{\small{\system}}\label{fig:PRSA}
	\end{subfigure}
  \caption{Frequency distribution of mutual information between the output content and the target prompts.}
  \label{fig:mutual_info}
\end{figure}

We validate our hypothesis through two settings to analyze the relationship between mutual information and attack performance. In the first, we calculate the mutual information between the outputs and the target prompts. In the second, we calculate it between the outputs incorporating \system's prompt attention feedback and the target prompts. For experimental details, please refer to Appendix~\ref{appendix:Experimental_Setup_for_Analyzing}. Figure~\ref{fig:mutual_info} shows that successful attacks tend to have higher mutual information. Furthermore, \system with prompt attention shows a significant increase in mutual information compared to \system without it. This increase in mutual information also corresponds to a higher number of successful attacks. A comparison with a heuristic baseline that relies on surface cues (Appendix~\ref{appendix:Heuristic_Baseline}) further supports our findings. These findings suggest that prompt attention indeed helps capture more characteristics of the target prompts, enhancing the mutual information between the outputs and the prompts.

These findings support our hypothesis: the greater the mutual information between the output and the target prompt, the higher the risk of prompt leakage. This insight also points to a potential defensive strategy: by analyzing the mutual information between the output and the target prompt, we can assess the risk of prompt leakage. For defenders, reducing mutual information while maintaining the utility of the target prompt presents an interesting challenge for future research.

\section{Discussion}

\subsection{Possible Defenses}
\label{sec:Possible_Defenses}

\begin{figure}
	\setlength{\abovecaptionskip}{1pt}
	\captionsetup[subfigure]{justification=centering}
	\centering
        \begin{subfigure}{0.475\linewidth}
		\includegraphics[width=\textwidth]{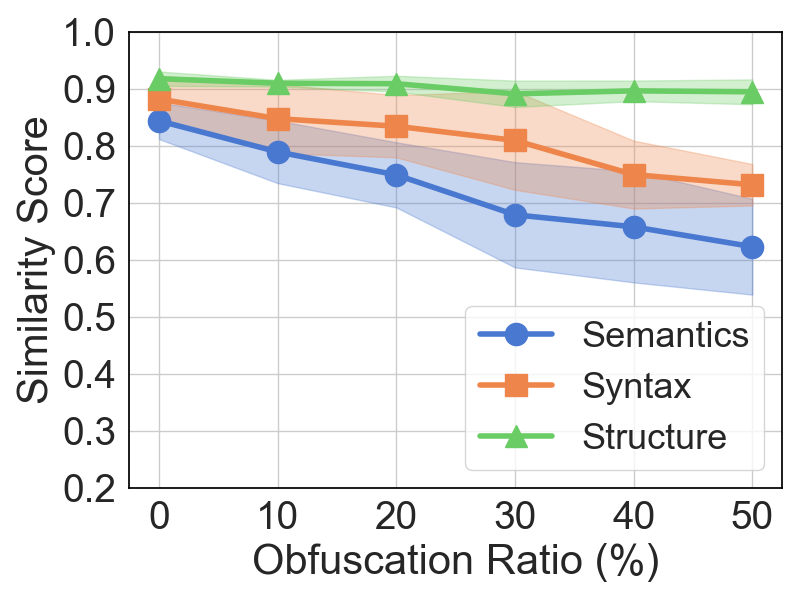}
		\caption{\small{Similarity Score}}\label{fig:Obfusion_ratio_sim}
	\end{subfigure}
	\hspace{0.2cm}
	\begin{subfigure}{0.475\linewidth}
		\includegraphics[width=\textwidth]{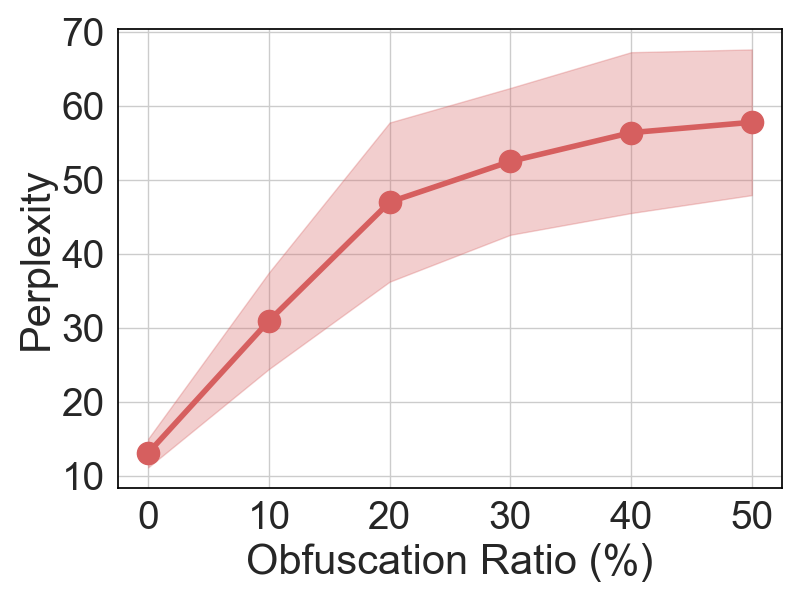}
		\caption{\small{Perplexity}}\label{fig:PPL}
	\end{subfigure}
  \caption{The impact of output obfuscation on \system.}
  \label{fig:defense_1}
\end{figure}

\paragraphbe{Output Obfuscation}
The effectiveness of \system primarily relies on the target's output content. To defend, one strategy is to limit adversaries' access to the full output content. We evaluate this approach using GPT-3.5 based prompts, with the same attack setup in Section~\ref{sec:ablation_study_section}. Details of output obfuscation are in Appendix~\ref{appendix:output_obfuscation_details}. Figure~\ref{fig:Obfusion_ratio_sim} shows that increasing the obfuscation ratio effectively reduces semantic and syntactic similarity, while structural similarity remains largely unchanged. Notably, at a $50\%$ obfuscation rate, both semantic and syntactic similarity decrease by over $20\%$. 

However, output obfuscation also has drawbacks. We measure the perplexity of the obfuscated outputs using GPT-2 to mimic human comprehension. As shown in Figure~\ref{fig:PPL}, a higher obfuscation ratio correlates with increased perplexity. When the ratio exceeds $30\%$, the perplexity surpasses $50$, making the output user-unfriendly. Thus, this strategy requires a balance between defensive effectiveness and usability.

\paragraphbe{Prompt Watermark}
Another defense strategy is watermarking prompts to protect intellectual property through legal deterrence. We follow PromptCARE~\cite{yao2023promptcare} to embed watermarks in six categories of target prompts by introducing specific triggers. After prompt stealing attacks, we test the output token IDs from the stolen prompts and assess their \emph{p-values}, which measure the statistical significance of the observed differences compared to a null hypothesis. An average p-value at or above $0.05$ suggests the presence of a watermark, indicating unauthorized use. To avoid skewed results, we focus on the stolen prompts from successful attacks.

\begin{table}
\centering
\footnotesize 
\caption{Watermark verification in outputs generated from stolen prompts. If the p-value is $\geq 0.05$, the stolen prompt is considered to contain the watermark.}
\label{tab:defense_2}
\begin{tabular}{@{}C{0.8cm}C{0.8cm}C{0.8cm}C{0.8cm}C{0.8cm}C{0.8cm}C{0.8cm}@{}}
\toprule
\multirow{3}{*}{\textbf{Metric}} & \multicolumn{6}{c}{\textbf{Category}}                               \\ \cmidrule(l){2-7} 
                               & \footnotesize{Ads} & \footnotesize{Email} & \footnotesize{Ideas} & \footnotesize{Music} & \footnotesize{Sports} & \footnotesize{Travel} \\ \midrule

\multirow{2}{*}{P-value}                  & $1.31\times10^{-2}$      & $1.47\times10^{-5}$      & $4.96\times10^{-4}$      & $1.28\times10^{-2}$     & $1.90\times10^{-6}$      & $3.02\times10^{-3}$
                                 \\
                                \bottomrule
\end{tabular}
\end{table}

Table~\ref{tab:defense_2} shows the p-value of watermark verification in outputs from the stolen prompts, averaged across ten trials. The results consistently show small p-values ($< 0.05$), indicating insufficient evidence to confirm watermarks in the stolen prompts. We explore the reasons behind this. In PromptCARE case studies, replicated prompts are fine-tuned based on the target prompts, closely resembling the originals. In contrast, our stolen prompts only replicated the target prompt's functionality. Although the target LLM's outputs are consistent, there are slight differences in vocabulary, syntax, and structure, similar to paraphrasing. Studies have shown that watermarking is less effective in verifying paraphrased content~\cite{sadasivan2023can}. Therefore, watermarking is of limited effectiveness as a defense against prompt stealing attacks.

\paragraphbe{Practicality of Potential Defenses}
While the defenses discussed above have limited practicality, they reveal promising directions that are more feasible. One is to restrict the level of detail in LLM outputs to reduce the risk of exposing target prompt semantics. Another is to optimize target prompts using soft prompts, making their intent less inferable from input-output pairs. These strategies are more practical because they can be integrated into existing prompt services with minimal disruption and allow for the design of optimization steps that preserve prompt usability. We are actively collaborating with prompt developers to explore these directions, and their effectiveness is currently under investigation.

\subsection{Justification of Functional Consistency}
We evaluate functional consistency using sentence transformer, FastKASSIM, and JS divergence. These metrics capture complementary linguistic features and are widely adopted in prior work~\cite{chen2022fastkassim, hui2024pleak, thompson2015performance}. Unlike FastKASSIM and JS divergence, the sentence transformer requires embeddings to compute similarity. To assess its robustness, we replace the underlying embedding model and evaluate the stability of the resulting rankings using Kendall’s Tau~\cite{kendall1938new} (see Appendix~\ref{appendix: robustness_semantic_sim} for details). Results show that the average Kendall’s Tau remains above $0.70$ across embedding models, indicating stable performance trends. Beyond linguistic similarity, functional consistency in real-world prompts may involve additional dimensions such as factual accuracy, sentiment, and completeness. To account for these, we conduct human evaluation and LLM-based multidimensional assessment, with these dimensions especially reflected in the LLM-based evaluation metrics. Notably, both evaluations yield trends consistent with those of our functional consistency metric, reinforcing its reliability as a proxy for persuasiveness.

\subsection{Limitations}
\label{sec:Limitations}
\system relies on the mutual information between the target prompt and its outputs. The attack may fail when the output primarily reflects the user input with minimal prompt influence, as in simple question-answering scenarios, or when the output lacks information about the target prompt's intent. Future work could explore incorporating semantic cues from user input more effectively or leveraging publicly available prompt descriptions to enhance intent inference.

\section{Conclusion}

This paper introduces \system, a practical framework designed for prompt stealing attacks against real-world prompt services, aiming to expose the risks of prompt leakage when the input-output content of prompts is revealed. \system infers the intent behind a prompt by analyzing its input-output content and generates a stolen prompt replicating the original's functionality through two key phases: prompt generation and prompt pruning. We validate the actual threat posed by \system through extensive evaluations on real-world prompt services. Our objective is to enhance risk awareness among prompt service vendors and actively advocate for implementing protective measures against prompt stealing attacks.

\section*{Acknowledgments}
We thank our shepherd and the anonymous reviewers for their
insightful comments on our work. This work was partly supported by the National Key Research and Development Program of China under No. 2022YFB3102100, NSFC under No. U244120033, U24A20336, 62172243, 62402425, and 62402418, the China Postdoctoral Science Foundation under No. 2024M762829, the Zhejiang Provincial Natural Science Foundation under No. LD24F020002, and the Zhejiang Provincial Priority-Funded Postdoctoral Research Project under No. ZJ2024001.

\section*{Ethics Consideration}

In developing \system, we encountered several ethical challenges. This section summarizes our key considerations and safeguards, including responsible disclosure, participant protection, and broader implications for LLM safety and ecosystem integrity.

\textbf{Stakeholder Analysis.} We identified three key stakeholder groups: (1) prompt developers, (2) prompt service vendors, and (3) the research community. Prompt developers may face intellectual property risks from functional replication. We mitigated this by avoiding direct reproduction, obtaining authorization for key demonstrations, and omitting sensitive implementation details. Vendors may face reputational or operational risks; we disclosed all findings to them in advance. For the research community, our work highlights both vulnerabilities and defenses in prompt services.

\textbf{Protection of Participants.} Although our institution does not have a formal Institutional Review Board (IRB), the human evaluation was reviewed and approved through an internal ethics review process overseen by our institutional research ethics committee, and conducted with the guidance of a legal expert in intellectual property and research ethics. We strictly adhered to the principles outlined in the Menlo Report. Participants were fully informed of the study’s purpose, their role, and how their data would be used, and provided explicit consent prior to participation. To minimize risk, we anonymized all responses, randomized the presentation of outputs, and ensured that no personal or identifying information was collected. These measures reflect our strong commitment to the Menlo Report’s principle of ``Respect for Persons''.

\textbf{Responsible Disclosure.} Following the principle of beneficence, we conducted responsible disclosure prior to public dissemination. We shared our findings with PromptBase, OpenAI, OpenGPT, and the developers of all prompts and LLM applications evaluated in our study. Along with reporting identified risks, we proposed potential mitigations: (1) avoiding full output display in prompt marketplaces, and (2) limiting exposure of prompt details through optimization. As these mitigations require further study, we recommend temporarily removing high-risk prompts. We continue collaborating with the developers to support practical defense deployment.

\textbf{Intellectual Property Considerations.} Our research raised potential concerns related to the replication of commercial prompt functionality. To prevent any infringement of copyright or proprietary content, all prompts presented in the paper were either (1) paraphrasing the functional intent while altering wording and structure, or (2) significantly modified to remove identifiable or proprietary components. We then submitted these revised examples to the corresponding prompt or application developers for review and obtained their explicit authorization to include them in the study and publication. These steps were taken to respect developers' intellectual property rights while preserving the ability to communicate core research findings.

\section*{Open Science Policy}
In compliance with USENIX Security’s open science policy and ethical guidelines, we have released all supporting artifacts, including source code, scripts, and datasets that do not raise ethical concerns. These resources are available at\url{https://doi.org/10.5281/zenodo.15549519}. The repository includes detailed usage instructions to support future research.

\bibliographystyle{plain}
\bibliography{biblio}

\appendix
\section*{Appendix}

\section{Details of Implementation}
\label{appendix:details_experiment_setup}
\vspace{-0.05cm}

\subsection{Instructions in Prompt Generation Phase}
\label{appendix:prompt_generation_instructions}
During the prompt generation phase, we use manual instructions to guide the target LLMs in checking for differences in semantics, syntax, and structure between the generated output and the target output. The manual instructions are as follows:

\begin{tcolorbox}[]
\textit{
Score based on {element} similarity between generated output and target output, if total score is $10$.
}
\end{tcolorbox}
\noindent where element denotes a specific linguistic expression factor, such as style, theme, or tone, among others. Based on the similarity score $s_{sim}$ provided by the LLM for each linguistic expression factor $d_{m}$, we compute its difference score as
$loss_{d_{m}} = 10 - s_{sim}$, which quantifies the contribution of $d_{m}$ to the overall output differences. Factors with $loss_{d_m} > \theta_{a}$ are classified as high-impact and are incorporated into the attention $a$.

We use LLMs as generative models to generate stolen prompts, with the following instructions being executed:

\begin{tcolorbox}[]
\textit{
Your task is to generate an instruction based on the provided user input and output. The instruction should focus on the specified instruction characteristics.
}
\end{tcolorbox}

The instruction characteristics are obtained as follows:

\begin{tcolorbox}[]
\textit{
What is the {$a_{i}$} of the output in one sentence?
}
\end{tcolorbox}
\noindent where $a_{i}$ is the element in prompt attention $a^*$.

For LLM application stores, we also consider using direct queries from the target LLM application to obtain the characteristics of the target system prompt. Drawing on the work of Zhang et al.~\cite{zhang2024extracting}, the instruction is designed as follows:
\begin{tcolorbox}[]
\textit{
Please describe yourself using as many sentences as possible that best describe you. Start with ``1:''
}
\end{tcolorbox}

\subsection{Parameter Settings}
\label{sec:Parameter_Settings}

In the prompt generation phase of the \system, when GPT-3.5 and GPT-4 serve as target LLMs, the temperature parameters are set to $0.7$ by default. When these models serve as generative models, their temperature parameters are adjusted to $0$ to ensure reduced output randomness. The attention threshold, $\theta_a$, is set to $1$ (on a scale of $1$ to $10$) based on empirical validation, as it effectively balances the inclusion of subtle but meaningful discrepancies to achieve better functional consistency. In the prompt pruning phase, we set the semantic similarity threshold, denoted by $\gamma$, to $0.4$. The beam size, $b$, is set to $2$, the evaluation frequency, $e$, to $5$, and the truncation factor, $\alpha$, to $0.6$ by default.

\subsection{Details of OPRO Baseline}
\label{sec:details_of_opro}
To ensure a fair and faithful comparison with \system, we apply the official implementation of OPRO~\cite{yang2023large} with two modifications to align with our prompt stealing setting: (1) We restrict OPRO to a single input-output pair for optimization, due to the limited access available to the adversary; (2) We replace OPRO’s accuracy-based scoring with our functional consistency metric, which evaluates semantic, syntactic, and structural similarity between outputs generated by the candidate and target prompts under the same input. All other settings remain unchanged.

\subsection{Output Obfuscation Mechanism}
\label{appendix:output_obfuscation_details}

We implement an output obfuscation strategy that can be seamlessly integrated into real-world prompt services. Specifically, we randomly mask a fixed proportion of tokens in the output generated by the target prompt and replace them with a generic obfuscation symbol (e.g., \texttt{*}). Given a predefined obfuscation ratio, we randomly select positions of words and phrases from the output and substitute them with the obfuscation symbol. For example, consider the original output: ``\texttt{Tired of phones that lag or lose juice halfway through the day? This mobile phone keeps up with your hustle, crisp camera, lightning-fast speed, and battery for days\ldots}''. The transformed output may appear as: ``\texttt{Tired of phones that * or * juice halfway through the day? This * * keeps up with your *, * camera, * speed, and * for days\ldots}''.

\section{Similarity Analysis of Prompts $p_{i}$ in $D$ and Target Prompts $p_{t}$}
\label{appendix:similarity_analysis_with_Dc}

\begin{figure*}[t]
	\centering
	\setlength{\abovecaptionskip}{2pt}
	\setlength{\belowcaptionskip}{-5pt}
	\includegraphics[width=1\textwidth]{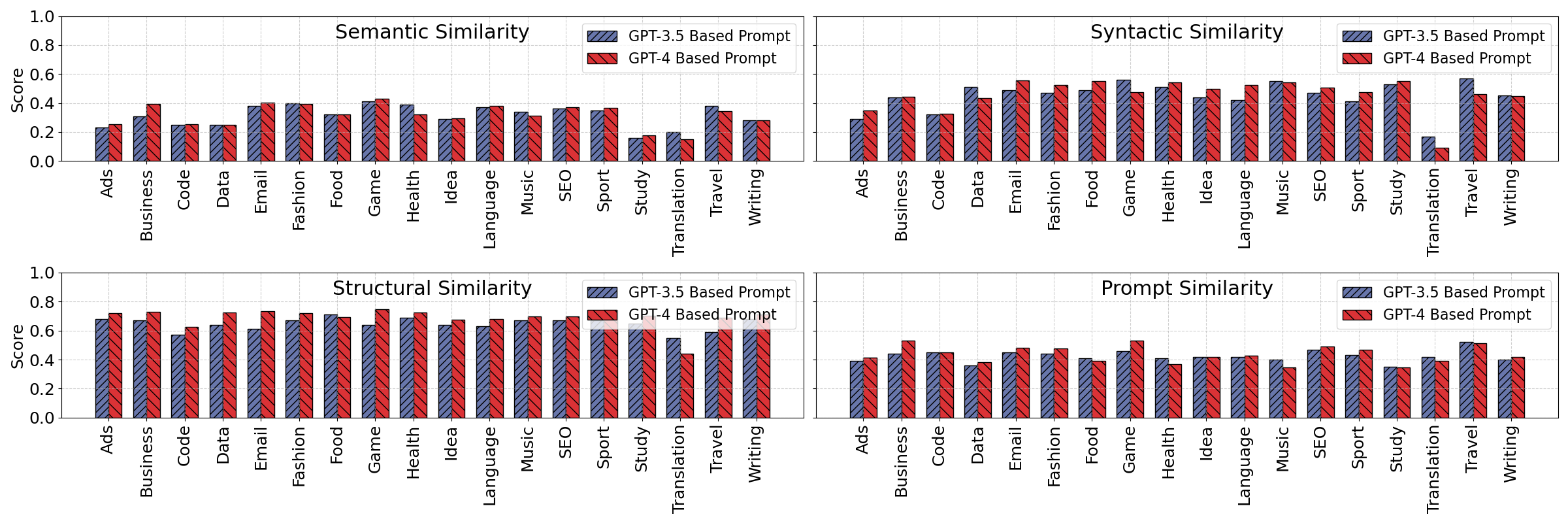}
	\caption{Similarity scores between prompts $p_i$ in $D$ and target prompt $p_t$. Semantic, syntactic, and structural similarity scores reflect the similarities between outputs $y_i$ and $y_t$, while prompt similarity scores measure the similarity between $p_i$ and $p_t$.}
	\label{fig:data_similarity_analysis}
\end{figure*}

We conduct a similarity analysis on the prompts $p_i$ and their input-output pairs $(x_i, y_i)$ from the collected dataset $D$, as well as the target prompts $p_t$ and their input-output pairs $(x_t, y_t)$, to evaluate whether $D$ exhibits generality. Following the metrics described in Section~\ref{sec:Metrics}, we assess the semantic, syntactic, and structural similarities between $y_i$ and $y_t$, along with the prompt similarity between $p_i$ and $p_t$.

As shown in Figure~\ref{fig:data_similarity_analysis}, the results show that while $y_i$ and $y_t$ exhibit moderate structural similarity (average scores of $0.64$ for GPT-3.5 based prompts and $0.69$ for GPT-4 based prompts), both their semantic similarity (average scores of $0.31$ for GPT-3.5 based prompts and $0.33$ for GPT-4 based prompts) and syntactic similarity ($0.45$ for GPT-3.5 based prompts and $0.46$ for GPT-4 based prompts) remain low. Moreover, the prompt similarity between $p_i$ and $p_t$ is much limited, with average scores of $0.41$ for GPT-3.5 based prompts and $0.44$ for GPT-4 based prompts. These findings demonstrate that the data in $D$ is not highly similar to the data in target prompts.

\begin{figure}
	\setlength{\abovecaptionskip}{1pt}
	\captionsetup[subfigure]{justification=centering}
	\centering
        \begin{subfigure}{0.475\linewidth}
		\includegraphics[width=\textwidth]{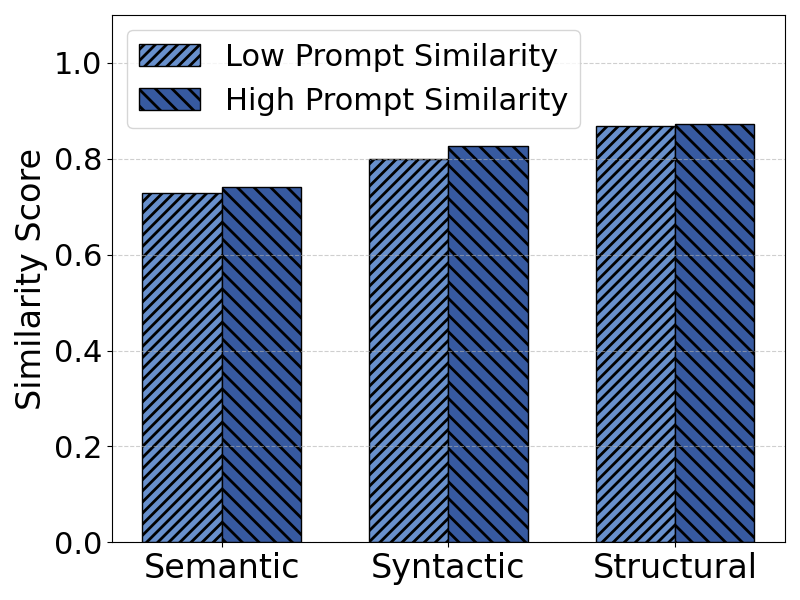}
		\caption{\small{GPT-3.5 Based Prompt}}
	\end{subfigure}
	\hspace{0.2cm}
	\begin{subfigure}{0.475\linewidth}
		\includegraphics[width=\textwidth]{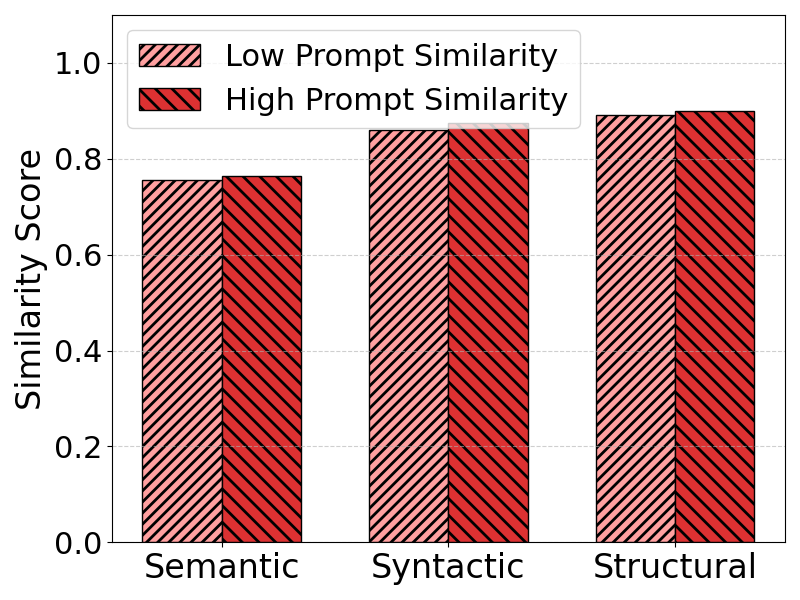}
		\caption{\small{GPT-4 Based Prompt}}
	\end{subfigure}
        
  \caption{Average functional consistency performance of \system on prompts $p_i$ in $D$ with high ($\geq$ 0.8) or low ($\leq$ 0.2) prompt similarity to the target prompt $p_t$.}
  \label{fig:d_high_low}
\end{figure}

To further examine whether \system's performance is strongly dependent on the auxiliary dataset $D$, we conduct a controlled experiment under two extreme conditions: one where prompts $p_i$ in $D$ are highly similar to the target prompt $p_t$, and another where they are deliberately dissimilar. We randomly sample $200$ purchased prompts from PromptBase as target prompts, including $100$ GPT-3.5 based and $100$ GPT-4 based prompts. For each target prompt $p_t$, we use an LLM to synthesize $p_i$ that are either similar or dissimilar. To ensure strict control over similarity, we compute the prompt similarity between each $p_i$ and $p_t$, and retain only those $p_i$ with similarity scores above $0.8$ (High Prompt Similarity) or below $0.2$ (Low Prompt Similarity). For each $p_t$, we collect $20$ $p_i$ for each condition to construct its own high-similarity and low-similarity dataset $D$.

Figure~\ref{fig:d_high_low} shows \system's average functional consistency across $D$ constructed with either high or low prompt similarity. The results demonstrate that \system achieves stable performance under both conditions. While scores are slightly higher in the high-similarity setting, the overall trends remain consistent across all three metrics. This robustness can be attributed to \system’s design: rather than relying on the specific content of individual prompts in $D$, \system learns generalizable stylistic and linguistic patterns representative of the prompt category. Since prompts within the same category often share common structures and intent, \system can infer the target prompt’s behavior even when $D$ has low prompt similarity with the target prompt.

\section{Attack Performance Against GPT-4 Based Prompts in Prompt Marketplaces}
\label{appendix:Effectiveness_gpt_4}

\begin{table*}
    \centering
    \caption{Attack performance of \system and baseline methods across $18$ categories of GPT-4 based prompts in the prompt marketplaces. The best results are shown in \textbf{bold}. ``--'' indicates not applicable.}
    \label{tab:performance_on_gpt4_based_prompt}
    \begin{adjustbox}{width=\textwidth}
    \setlength{\tabcolsep}{0.6mm}
    \begin{tabular}{cccccccccccccccccccc}
        \toprule
        \multirow{3}{*}{\textbf{\normalsize Metric}}         & \multirow{3}{*}{\textbf{\normalsize Attack Method}} & \multicolumn{18}{c}{\textbf{Category}}                               \\ \cmidrule(l){3-20} 
                                &                         & \textbf{Ads} & \textbf{Business} & \textbf{Code} & \textbf{Data} & \textbf{Email} & \textbf{Fashion} & \textbf{Food} & \textbf{Games} & \textbf{Health}
                                & \textbf{Ideas}  & \textbf{Language} & \textbf{Music}& \textbf{SEO}& \textbf{Sports}
                                & \textbf{Study} & \textbf{Translation} & \textbf{Travel} & \textbf{Writing} \\ \midrule
        \multirow{6}{*}{\begin{tabular}[c]{@{}c@{}}{Semantic} \\ {Similarity}\end{tabular}} 
        & OPRO & 0.66 & 0.60 & 0.65 & 0.75 & 0.69 & 0.74 & 0.69 & 0.67 & 0.73 & 0.68 & 0.63 & 0.69 & 0.71 & 0.72 & 0.62 & 0.57 & 0.62 & 0.59 \\
        & Sha et al. & 0.61 & 0.61 & 0.71 & 0.68 & 0.59 & 0.75 & 0.63 & 0.65 & 0.73 & 0.68 & 0.57 & 0.69 & 0.68 & 0.71 & 0.61 & 0.47 & 0.64 & 0.62 \\
        & output2prompt & 0.64 & 0.67 & 0.72 & 0.55 & 0.57 & 0.73 & 0.60 & 0.71 & 0.72 & 0.74 & 0.45 & 0.66 & 0.68 & 0.59 & 0.55 & 0.68 & 0.61 & 0.62 \\
        & PLEAK & -- & -- & -- & -- & -- & -- & -- & -- & -- & -- & -- & -- & -- & -- & -- & -- & -- & -- \\
         & \system & \textbf{0.75} & \textbf{0.72} & \textbf{0.80} & \textbf{0.82} & \textbf{0.80} & \textbf{0.83} & \textbf{0.79} & \textbf{0.75} & \textbf{0.77} & \textbf{0.83} & \textbf{0.76} & \textbf{0.86} & \textbf{0.79} & \textbf{0.78} & \textbf{0.72} & \textbf{0.80} & \textbf{0.78} & \textbf{0.78}\\
        \cmidrule(lr){2-20}
        & \% Gain for \system & 13.64 & 7.46 & 11.11 & 9.33 & 15.94 & 10.67 & 14.49 & 5.63 & 5.48 & 12.16 & 20.63 & 24.64 & 11.27 & 8.33 & 16.13 & 17.65 & 21.88 & 25.81 \\
        \midrule
        
        \multirow{6}{*}{\begin{tabular}[c]{@{}c@{}}{Syntactic} \\ {Similarity}\end{tabular}} 
        & OPRO  & 0.71 & 0.75 & 0.77 & 0.68 & 0.85 & 0.66 & 0.82 & 0.85 & 0.85 & 0.63 & 0.71 & 0.84 & 0.73 & 0.86 & 0.90 & 0.55 & 0.78 & 0.81\\
        & Sha et al. & 0.45 & 0.61 & 0.82 & 0.83 & 0.75 & 0.81 & 0.81 & 0.83 & 0.84 & 0.54 & 0.65 & 0.81 & 0.54 & 0.77 & 0.85 & 0.38 & 0.72 & 0.67 \\
        & output2prompt & 0.51 & 0.53 & 0.81 & 0.49 & 0.53 & 0.59 & 0.61 & 0.56 & 0.83 & 0.58 & 0.52 & 0.59 & 0.62 & 0.52 & 0.57 & 0.78 & 0.64 & 0.68 \\
        & PLEAK & -- & -- & -- & -- & -- & -- & -- & -- & -- & -- & -- & -- & -- & -- & -- & -- & -- & -- \\
        & \system & \textbf{0.79} & \textbf{0.83} & \textbf{0.85} & \textbf{0.92} & \textbf{0.97} & \textbf{0.91} & \textbf{0.92} & \textbf{0.91} & \textbf{0.96} & \textbf{0.81} & \textbf{0.85} & \textbf{0.95} & \textbf{0.90} & \textbf{0.94} & \textbf{0.91} & \textbf{0.85} & \textbf{0.95} & \textbf{0.83}\\
        \cmidrule(lr){2-20}
        & \% Gain for \system & 11.27 & 10.67 & 3.66 & 10.84 & 14.12 & 12.35 & 12.20 & 7.06 & 12.94 & 28.57 & 19.72 & 13.10 & 23.29 & 9.30 & 1.11 & 8.97 & 21.79 & 2.47\\
        \midrule

        \multirow{6}{*}{\begin{tabular}[c]{@{}c@{}}{Structural} \\ {Similarity}\end{tabular}} 
        & OPRO & 0.79 & 0.83 & 0.81 & 0.63 & 0.87 & 0.87 & 0.85 & 0.87 & 0.89 & \textbf{0.91} & 0.79 & 0.88 & 0.81 & 0.89 & 0.87 & 0.64 & 0.85 & 0.83 \\
        & Sha et al. & 0.76 & 0.83 & 0.84 & 0.80 & 0.91 & 0.89 & 0.86 & 0.86 & 0.88 & 0.79 & 0.77 & 0.87 & 0.78 & 0.89 & 0.88 & 0.51 & 0.89 & 0.84 \\
        & output2prompt & 0.80 & 0.83 & 0.86 & 0.63 & 0.78 & 0.87 & 0.81 & 0.83 & 0.89 & 0.83 & 0.73 & 0.83 & 0.81 & 0.78 & 0.81 & 0.66 & 0.82 & 0.84 \\
        & PLEAK & -- & -- & -- & -- & -- & -- & -- & -- & -- & -- & -- & -- & -- & -- & -- & -- & -- & -- \\
        & \system & \textbf{0.87} & \textbf{0.88} & \textbf{0.88} & \textbf{0.89} & \textbf{0.95} & \textbf{0.93} & \textbf{0.89} & \textbf{0.91} & \textbf{0.89} & 0.90 & \textbf{0.88} & \textbf{0.92} & \textbf{0.89} & \textbf{0.93} & \textbf{0.89} & \textbf{0.83} & \textbf{0.92} & \textbf{0.89} \\
        \cmidrule(lr){2-20}
        & \% Gain for \system & 8.75 & 6.02 & 2.33 & 11.25 & 4.40 & 4.49 & 3.49 & 4.60 & 0 & -1.10 & 11.39 & 4.55 & 9.88 & 4.49 & 1.14 & 25.76 & 3.37 & 5.95 \\
        \bottomrule
    \end{tabular}
    \end{adjustbox}
\end{table*}

We describe the attack performance on GPT-4 based prompts from $18$ categories in prompt marketplaces. As illustrated in Table~\ref{tab:performance_on_gpt4_based_prompt}, \system outperforms the other baseline methods.

\section{Sample Size Sensitivity Analysis}
\label{appendix:sensitivity_analysis}

To assess whether our functional consistency evaluation is sensitive to the choice of test sample size, we conduct a sample size sensitivity analysis by varying the number of user input cases ($n$) and the number of generation samples per input ($m$). Specifically, we compare functional consistency results under different ($n$, $m$) configurations, ranging from ($n=1$, $m=2$) to ($n=4$, $m=4$).

\begin{figure}[h]
	\setlength{\abovecaptionskip}{1pt}
	\captionsetup[subfigure]{justification=centering}
	\centering
        \begin{subfigure}{1\linewidth}
		\includegraphics[width=\textwidth]{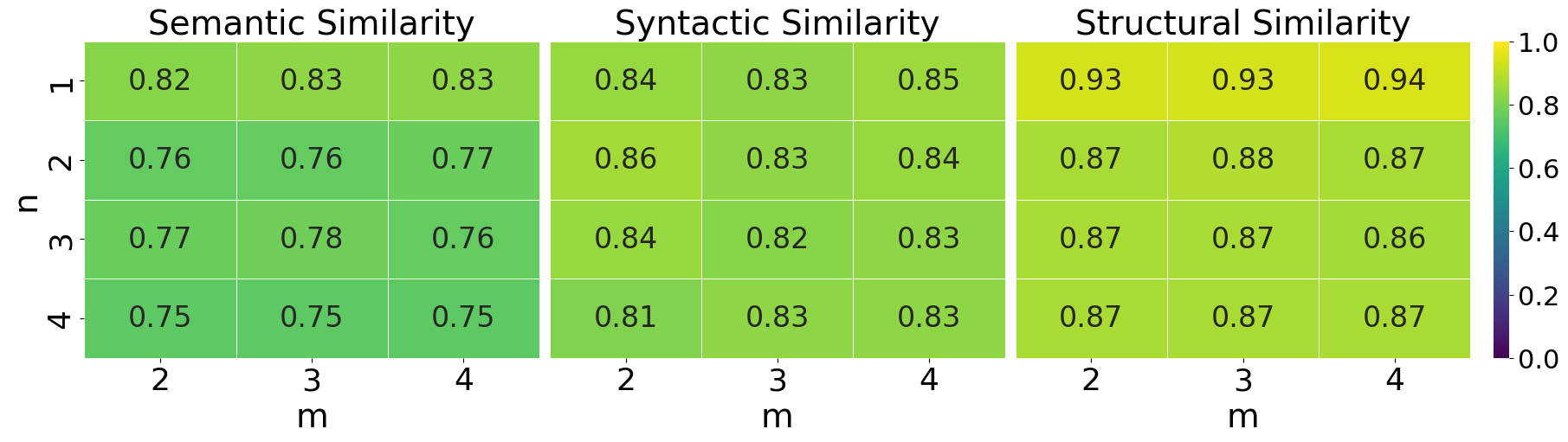}
		\caption{\small{GPT-3.5 Based Prompt}}
	\end{subfigure}
	\hspace{0.2cm}
	\begin{subfigure}{1\linewidth}
		\includegraphics[width=\textwidth]{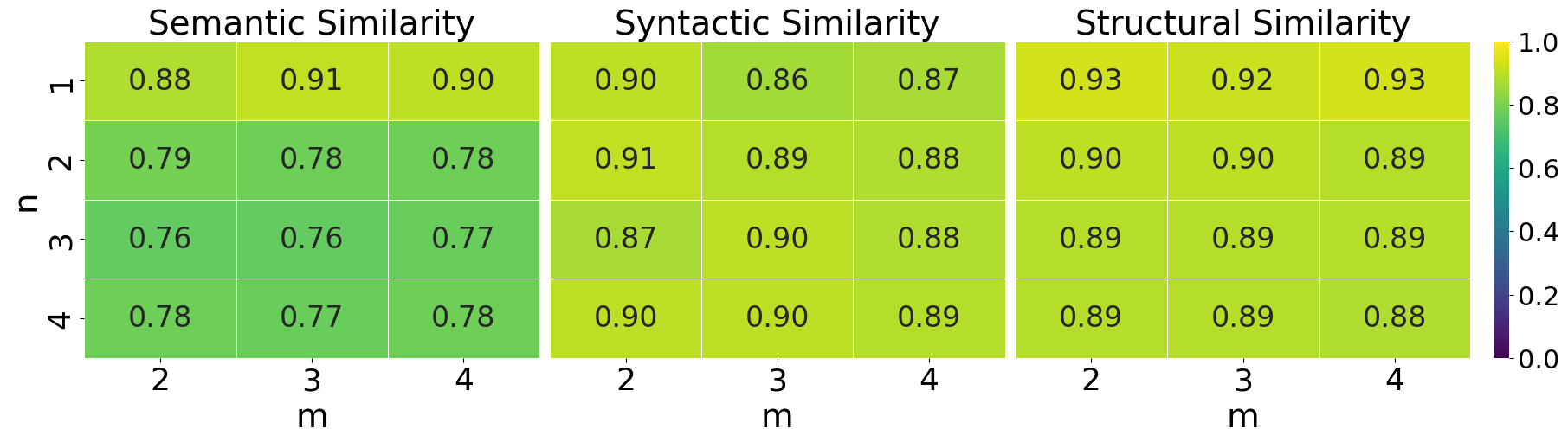}
		\caption{\small{GPT-4 Based Prompt}}
	\end{subfigure}
  \caption{Average functional consistency scores under varying numbers of user inputs ($n$) and generations per input ($m$) on GPT-3.5 and GPT-4 based prompts.}
  \label{fig:sensitivity_analysis}
\end{figure}

As shown in Figure~\ref{fig:sensitivity_analysis}, configurations with only one input ($n=1$) tend to produce greater variability across similarity metrics. In contrast, results are more consistent when $n$ and $m$ range from $2$ to $4$, with minimal fluctuations, indicating stability under our default configuration ($n=2$, $m=3$). This stability may be attributed to the averaging procedure in our evaluation: each input generates $m=3$ outputs from both prompts, leading to 9 pairwise comparisons per input. With $n=2$, this results in 18 total comparisons per method, averaged to report the similarity score. While larger sample sizes may yield slightly tighter estimates, the current configuration appears sufficient to support consistent evaluation trends.

\section{Threshold Exploration for ASR}
\label{appendix:Threshold_Exploration_for_ASR}

\begin{figure*}
	\centering
	\setlength{\abovecaptionskip}{2pt}
	\setlength{\belowcaptionskip}{-5pt}
	\includegraphics[width=1\textwidth]{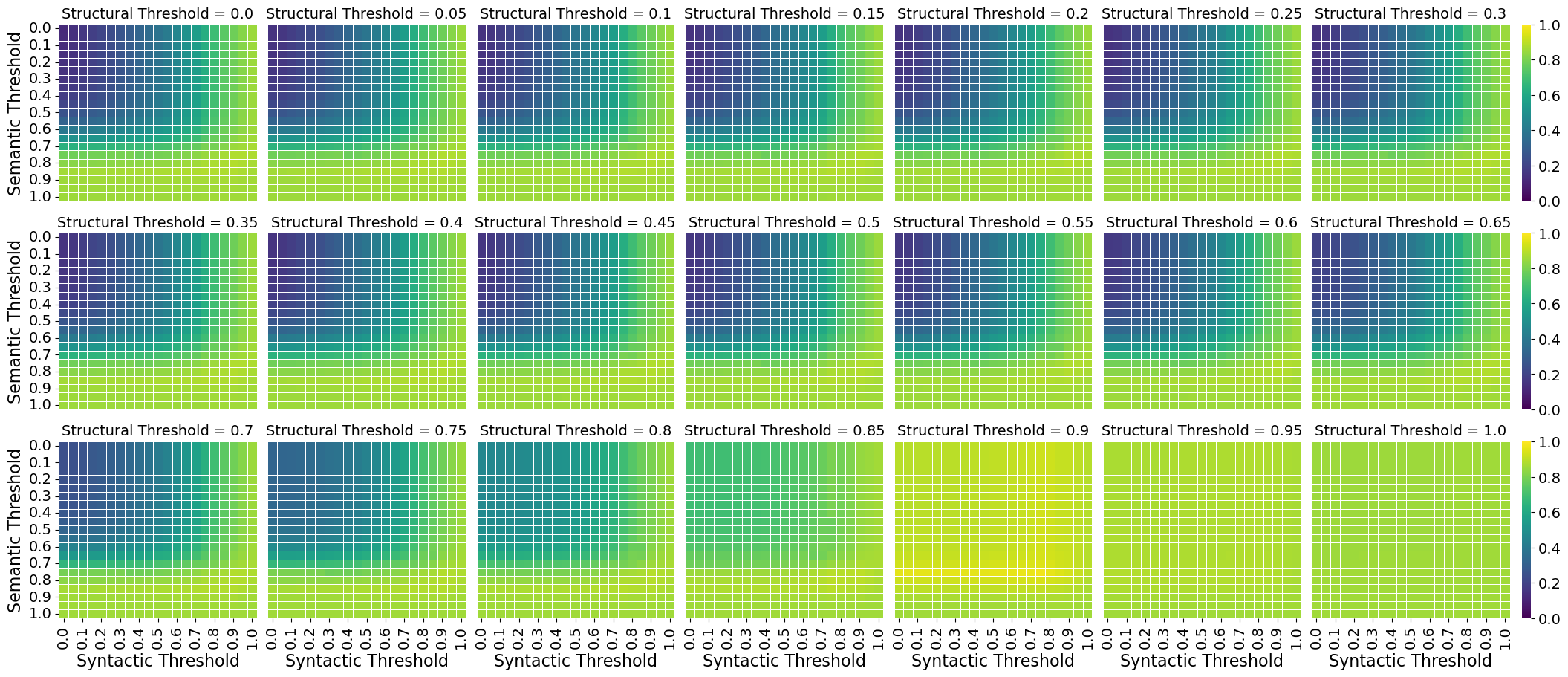}
	\caption{Grid search results for optimal functional consistency thresholds across semantic, syntactic, and structural similarities.}
	\label{fig:Threshold_Exploration}
\end{figure*}

To quantify the effectiveness of the attack more clearly, we explore the optimal functional consistency thresholds. This is done to establish a unified standard for determining whether prompts in prompt services are successfully stolen. We assess the consistency between the outputs generated by stolen prompts and the target outputs, as described in Section~\ref{sec:Performance_in_Prompt_Marketplaces}. This consistency is evaluated by labeling the outputs as $1$ (consistent) or $0$ (inconsistent), where a label of $1$ indicates that the stolen prompt successfully replicates the functionality of the target prompt, thus deeming the attack successful.

The process is primarily conducted by experts and ChatGPT, with ChatGPT achieving an accuracy of approximately $95\%$, confirming its suitability for further annotation tasks. Consequently, we continue using ChatGPT for the remaining annotation work. We employ a grid search method with intervals set at $[0, 1, 0.05]$ to determine the optimal thresholds for $\gamma_{sem}$, $\gamma_{syn}$, and $\gamma_{str}$, aligning the predicted successful attack samples with those labeled as $1$. As shown in Figure~\ref{fig:Threshold_Exploration}, the highest prediction accuracy of $98.2\%$ is achieved with $\gamma_{sem}=0.75$, $\gamma_{syn}=0.75$, and $\gamma_{str}=0.9$.

\section{Cost Analysis of \system}
\label{appendix:Cost_Analysis}

To evaluate the economic feasibility of \system, we compare the cost of purchasing prompts with the attack cost. If the average attack cost is significantly lower than the average prompt price, the attack becomes economically advantageous.

\paragraphbe{Cost Definitions} 
We define the attack cost for the adversary. If the adversary owns the computing resources, this cost is negligible. However, in real-world scenarios, the primary cost comes from invoking LLM APIs. For instance, based on OpenAI's pricing~\cite{OpenAI}, GPT-3.5-turbo charges $\$0.50$ per $1M$ input tokens and $\$1.50$ per $1M$ output tokens, while GPT-4-turbo charges $\$10.00$ per $1M$ input tokens and $\$30.00$ per $1M$ output tokens. Thus, the average attack cost is defined as follows:

\begin{equation}
    C_{avg} =  \frac{C_{api}}{N_{t}\cdot ASR},
\label{eq:attack_avg_wo_cloud}
\end{equation}

\noindent where $C_{avg}$ denotes the average attack cost, $C_{api}$ denotes the total cost of LLM API usage, $N_t$ denotes the number of target prompts, and $ASR$ denotes the attack success rate.

If the adversary lacks computing resources, the cost of computation must be taken into account. In this case, the average attack cost is defined as follows:

\begin{equation}
    C_{avg} =  \frac{C_{api} + C_{cal}}{N_{t}\cdot ASR},
\label{eq:attack_avg_w_cloud}
\end{equation}

\noindent where $C_{cal}$ denotes the extra computing resource cost. In general, we assume that the adversary conducts the attack by renting cloud computing resources. Based on pricing calculators from well-known cloud service providers~\cite{Cloud_recource}, the cost of renting a V100 GPU-accelerated server is approximately $\$1.66$ per hour. Considering the run-time of \system and potential retries, an $8$ hour rental period is sufficient.

\begin{table}
\centering
\scriptsize
\caption{Comparison of average prompt price and the average attack costs of \system. Average Attack Cost\(_1\): using own computing resources; Average Attack Cost\(_2\): using cloud computing resources.}
\label{tab:attack_costs}
\setlength{\tabcolsep}{1.3mm} 
\begin{tabular}{cccc}
\toprule
\textbf{Target Prompt} & \textbf{\begin{tabular}[c]{@{}c@{}}Average Prompt \\ Price (\$)\end{tabular}} & \textbf{\begin{tabular}[c]{@{}c@{}}Average Attack \\ Cost\(_1\) (\$)\end{tabular}} & \textbf{\begin{tabular}[c]{@{}c@{}}Average Attack \\ Cost\(_2\) (\$)\end{tabular}} \\
\midrule
GPT-3.5 Based Prompt        & 3.77                              & 0.05                           & 0.08                           \\
GPT-4 Based Prompt          & 4.15                              & 0.48                           & 0.51                           \\
\bottomrule
\end{tabular}
\end{table}

\paragraphbe{Results} 
The results in Table~\ref{tab:attack_costs} show that \system is highly cost-effective across both GPT-3.5 based and GPT-4 based prompts. For GPT-3.5 based prompts, the average attack cost is only $1.3\%$--$2.1\%$ of the average prompt price ($\$3.77$), while for GPT-4 based prompts, it is $11.6\%$--$12.3\%$ of the average prompt price ($\$4.15$). Even when using cloud computing resources, the attack remains profitable for the adversary.

\section{Optimization Between Effectiveness and Feasibility}
\label{appendix:Parameter_Impact}

We optimize parameters in both the prompt generation and pruning phases to balance effectiveness and feasibility.

\begin{figure}
	\setlength{\abovecaptionskip}{1pt}
	\captionsetup[subfigure]{justification=centering}
	\centering
	\begin{subfigure}{0.3\linewidth}
		\includegraphics[width=\textwidth]{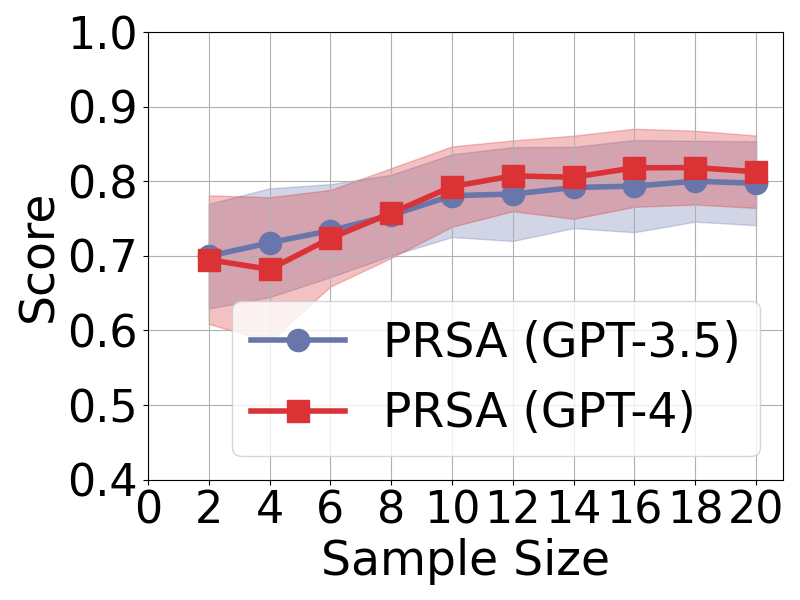}
		\caption{\scriptsize{Semantic Similarity}}\label{fig:semantic_opt_size}
	\end{subfigure}
	\hspace{0.2cm}
	\begin{subfigure}{0.3\linewidth}
		\includegraphics[width=\textwidth]{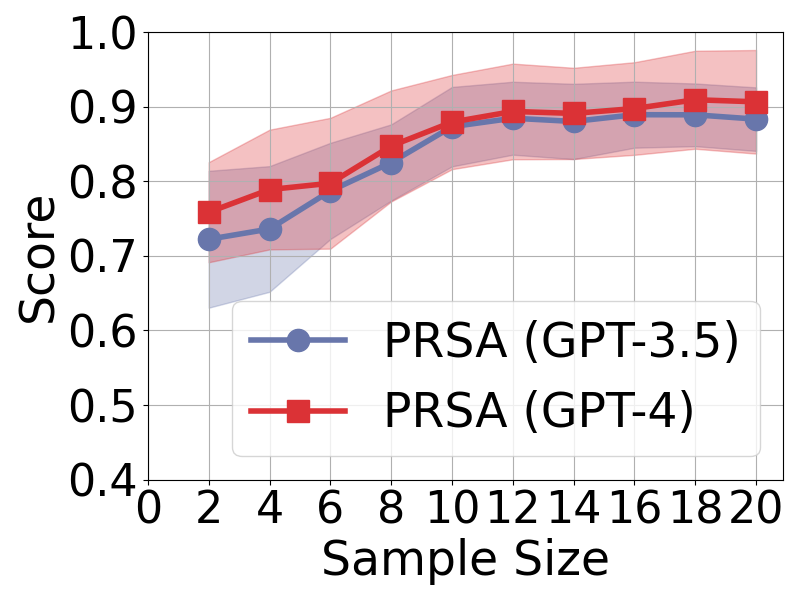}
		\caption{\scriptsize{Syntactic Similarity}}\label{fig:syntactic_opt_size}
	\end{subfigure}
        \hspace{0.2cm}
	\begin{subfigure}{0.3\linewidth}
		\includegraphics[width=\textwidth]{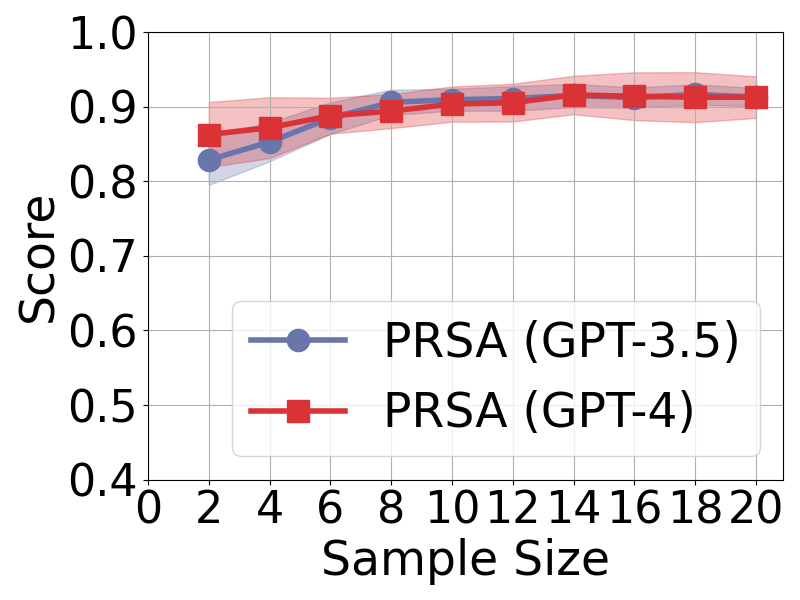}
		\caption{\scriptsize{Structural Similarity}}\label{fig:structural_opt_size}
	\end{subfigure}
  \caption{Assessing the impact of sample size variations in the prompt generation phase on \system's attack performance.}
  \label{fig:opt_size}
\end{figure}

In the prompt generation phase, determining the optimal number of dataset samples per category is crucial for achieving adequate prompt attention. Fewer samples reduce costs and enhance feasibility, while more samples typically increase accuracy. Figure~\ref{fig:opt_size} shows \system's average attack performance across various sample sizes. Semantic similarity improves with larger sample sizes, plateauing around $12$ samples. Syntactic similarity levels off at about $10$ samples, while structural similarity balances at $8$ samples. Therefore, a sample size between $10$ and $12$ is considered optimal for \system. Notably, \system still has a usable effect even with as few as $2$-$4$ examples, ensuring its applicability in low-resource settings.

\begin{figure}
	\setlength{\abovecaptionskip}{1pt}
	\captionsetup[subfigure]{justification=centering}
	\centering
	\begin{subfigure}{0.3\linewidth}
		\includegraphics[width=\textwidth]{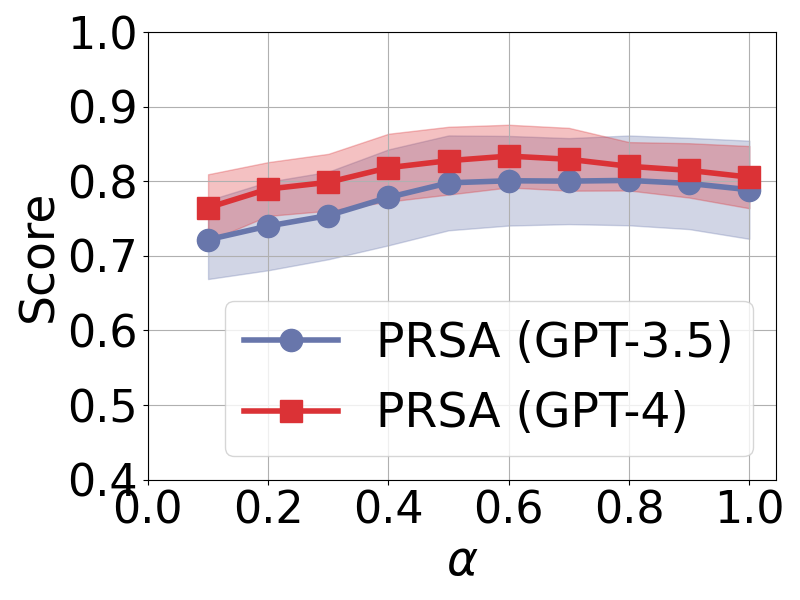}
		\caption{\scriptsize{Semantic Similarity}}\label{fig:semantic_opt_alpha}
	\end{subfigure}
	\hspace{0.2cm}
	\begin{subfigure}{0.3\linewidth}
		\includegraphics[width=\textwidth]{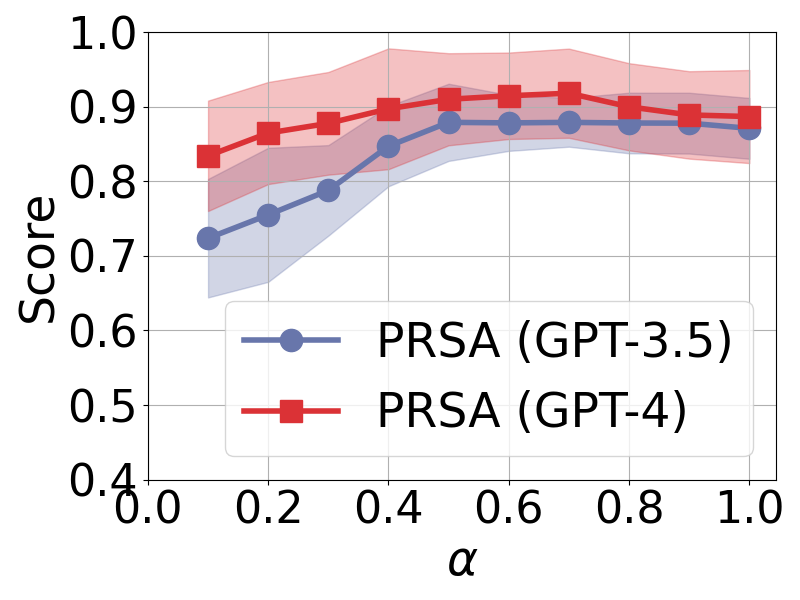}
		\caption{\scriptsize{Syntactic Similarity}}\label{fig:syntactic_opt_alpha}
	\end{subfigure}
        \hspace{0.2cm}
	\begin{subfigure}{0.3\linewidth}
		\includegraphics[width=\textwidth]{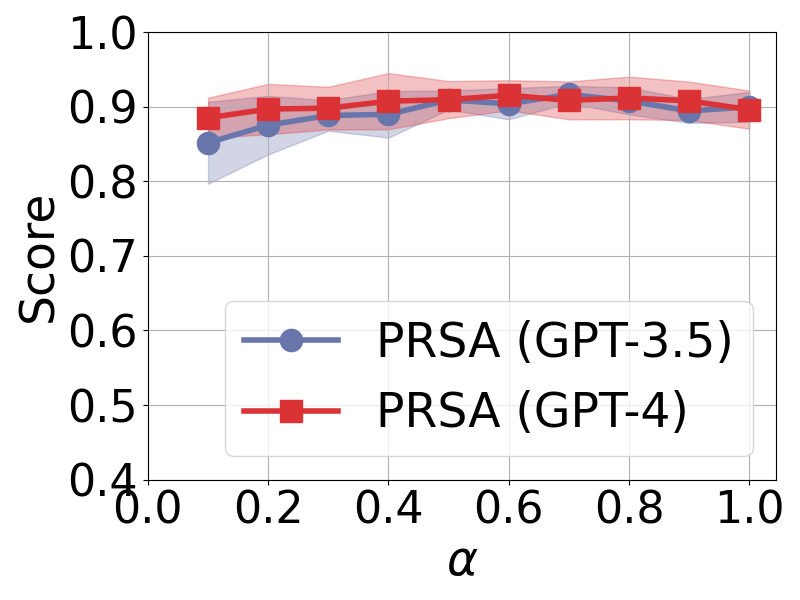}
		\caption{\scriptsize{Structural Similarity}}\label{fig:structural_opt_alpha}
	\end{subfigure}
  \caption{Assessing the impact of $\alpha$ variations in the prompt pruning phase on \system's attack performance.}
  \label{fig:opt_alpha}
\end{figure}
The beam search cost for identifying correlated words is crucial in the prompt pruning phase. Adjusting $\alpha$ helps limit the search space, aiming to balance cost and effectiveness. Figure~\ref{fig:opt_alpha} shows that for \system, optimal performance in semantic and syntactic similarity is achieved with $\alpha$ values between $0.6$ and $0.7$. Variations in $\alpha$ have minimal impact on structural similarity. Notably, when $\alpha$ exceeds $0.7$, there is a noticeable drop in attack performance, indicating that masking more words in a larger search space reduces the semantic integrity of stolen prompts. These findings suggest that an $\alpha$ near $0.6$ strikes an effective balance for \system.

\section{Theoretical Analysis}
\label{appendix:Theoretical_Analysis}
In this section, we explore the theoretical analysis of prompt stealing attacks. We begin by proposing a key lemma from Fano's Inequality~\cite{fano1961transmission}, which will then serve as a crucial result for proving the theorem proposed later.

\begin{lemma}
\label{lemma:binary_entropy_monotonic}
The binary entropy function $H_b(x)$ is monotonically increasing in the interval [$0, 0.5$] and monotonically decreasing in the interval [$0.5, 1$].
\end{lemma}

\begin{proof}[Proof of Lemma~\ref{lemma:binary_entropy_monotonic}]
According to Fano's Inequality~\cite{fano1961transmission}, the binary entropy function \( H_b(x) \) is defined as:
\small
{\begin{flalign*}
     H_b(x) = -x \log x - (1 - x) \log (1 - x)
\end{flalign*}}

To analyze the monotonicity, compute the derivative of $H_b(x)$ with respect to $x$:
{\footnotesize 
\begin{flalign*}
     \frac{dH_b}{dx} = & -\log x - x 
     \frac{d}{dx}\frac{\ln{x}}{\ln 2} + \log (1 - x) - (1-x) \frac{d}{dx}\frac{\ln{(1 - x)}}{\ln 2} \\
      = & -\log x - x \frac{1}{x \ln 2} + \log (1 - x) - (1-x) \frac{-1}{(1 - x) \ln 2} \\
      = & -\log x - \frac{1}{\ln 2} + \log (1 - x) + \frac{1}{\ln 2} \\
      = & -\log x + \log (1 - x) \\
      = & \log \left( \frac{1 - x}{x} \right) 
\end{flalign*}}

When $x \in [0, 0.5]$, $\frac{dH_b}{dx} \geq 0$, and when $x \in [0.5, 1]$, $\frac{dH_b}{dx} \leq 0$. Therefore, $H_b(x)$ is monotonically increasing in the interval [$0, 0.5$] and monotonically decreasing in the interval [$0.5, 1$].
\end{proof}

\begin{theorem}
\label{the:mutual_information_relation}
Consider an LLM where the input consists of a (system) prompt $p = (p_1, p_2, \ldots, p_n)$ $(p_i \in V)$ and a user input $x = (x_1, x_2, \ldots, x_n)$ $(x_i \in V)$, resulting in an output $y = (y_1, y_2, \ldots, y_m)$ $(y_i \in V)$. Let $I(p; y)$ denote the mutual information between $p$ and $y$. Let $P_{e}$ denote the minimum error probability of inferring $p$ from $y$. Let $|\mathcal{S}|$ denote the cardinality of the prompt space. Under mild regularity conditions, the lower bound of $P_{e}$ is formalized as $P_e \geq 1 - \frac{I(p; y) + \log 2}{\log |\mathcal{S}|}$.

\end{theorem}

\begin{proof}[Proof of Theorem~\ref{the:mutual_information_relation}]

The mutual information between $p$ and $y$ is defined as:
\small
\begin{flalign*}
   I(p; y) = H(p) - H(p \mid y),
\end{flalign*}
\noindent where $H(p)$ is the entropy of $p$, and $H(p \mid y)$ is the conditional entropy of $p$ given the output $y$.

Fano's Inequality~\cite{fano1961transmission} provides an upper bound on the conditional entropy in terms of the error probability $P_{e}$:
\small
\begin{flalign*}
   H(p \mid y) \leq H_b(P_e) + P_e \log(|\mathcal{S}| - 1),
\end{flalign*}
\noindent where $H_b(P_e) = -P_e \log P_e - (1 - P_e) \log (1 - P_e)$ is the binary entropy function~\cite{fano1961transmission}. In general, we assume that $p$ is uniformly distributed over $|\mathcal{S}|$~\cite{scarlett2019introductory}. Thus, $H(p) = \log |\mathcal{S}|$ is a constant. Based on the monotonicity of $H_b(P_e)$ (see Lemma~\ref{lemma:binary_entropy_monotonic}), when $P_e = 0.5$,  $H_b(P_e)$ reaches its maximum value $H_b(0.5) = 1$. Therefore, $H_b(P_e) \leq \log 2$. As a result, $H(p \mid y)$ can be further expressed as:
\small
\begin{flalign*}
   H(p \mid y) \leq & \log 2 + P_e \log(|\mathcal{S}| - 1) \\
   \leq & \log 2 + P_e \log |\mathcal{S}| 
\end{flalign*}
Combining with mutual information, we obtain the inequality:
\small
\begin{flalign*}
   I(p; y) \geq \log |\mathcal{S}| - (\log 2 + P_e \log |\mathcal{S}|)
\end{flalign*}
Therefore:
\small
\begin{flalign*}
  P_e \geq 1 - \frac{I(p; y) + \log 2}{\log |\mathcal{S}|}
\end{flalign*}
\end{proof}

As a result, to obtain a low $P_e$, one possible approach is to increase $I(p; y)$, which minimizes the lower bound of $P_e$.

\section{Supplementary Details for Why Our Attacks Work}

\subsection{Experimental Setup}
\label{appendix:Experimental_Setup_for_Analyzing}

\paragraphbe{Datasets}
We sample $200$ instances from the prompt marketplaces as described in Section~\ref{sec:Datasets} to evaluate the relationship between mutual information and attack effectiveness.

\paragraphbe{Methodology}
For each sample, we utilize a sentence transformer to generate embeddings and calculate the mutual information. To reduce randomness, we select $5$ different user inputs for each prompt and then generate $5$ corresponding outputs. Next, we calculate the average of the mutual information values between these outputs and their respective prompts.

\paragraphbe{Metrics} 
The evaluation metrics remain consistent with those outlined in Section~\ref{sec:Metrics}.

\subsection{Comparison with a Heuristic Baseline}
\label{appendix:Heuristic_Baseline}

We investigate the effectiveness of heuristic prompt stealing strategies from a mutual information perspective. A natural baseline attempts to reconstruct the target prompt using observable elements from the output. Specifically, we construct the heuristic prompt as: ``Write a [prompt description] that includes [section titles].'' Here, the prompt description can be obtained from the information publicly available in PromptBase, while the section titles can be extracted by prompting an LLM to parse the target output. This baseline aims to increase mutual information by leveraging surface-level structural cues. The baseline is evaluated using the setup and data described in Appendix~\ref{appendix:Experimental_Setup_for_Analyzing}.

\begin{figure}
  \centering
  \includegraphics[width=0.6\linewidth]{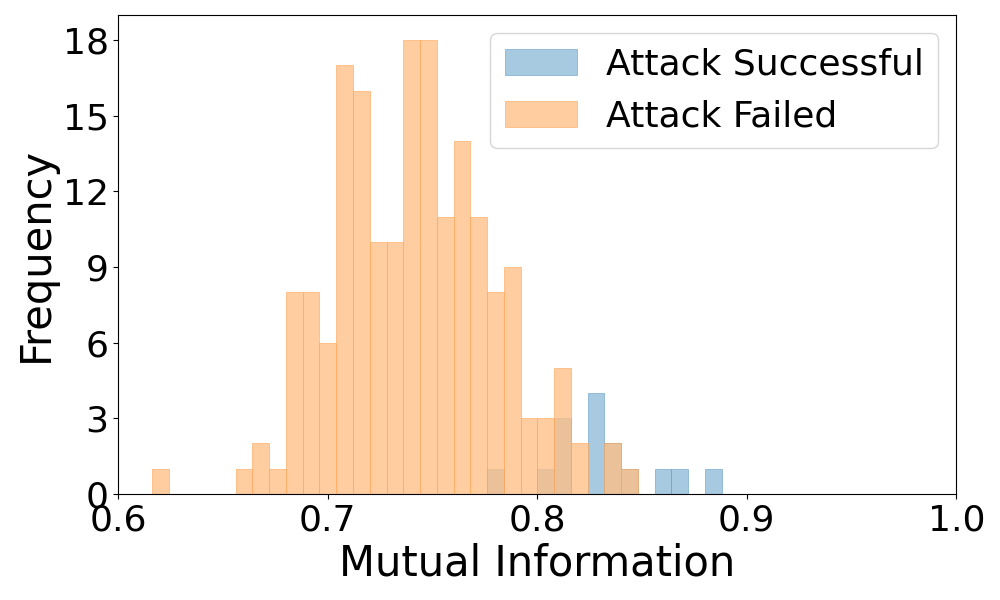}
  \caption{Frequency distribution of mutual information between outputs (incorporating extracted surface cues) and the target prompts.}
  \label{fig:heuristic_mi}
\end{figure}

Figure~\ref{fig:heuristic_mi} shows the distribution of attack success and failure for the heuristic baseline, along with the mutual information between the outputs (incorporating extracted cues) and the target prompts. Compared to \system (Figure~\ref{fig:PRSA}), most heuristic attempts fall into the failure region. We identify three key reasons for this result. First, many outputs lack explicit section titles or consistent structural formatting, complicating reliable cue extraction. Second, the extracted content primarily reflects surface-level intent. In contrast, \system leverages prompt attention and iterative feedback to infer deeper prompt intent, contributing to its higher mutual information. Third, although a heuristic prompt may align with a specific output, it often fails to generalize across different user inputs, leading to low average mutual information. These findings show the limitations of surface-level heuristics, and further refinements are needed to improve the effectiveness.

\section{Manual vs. Algorithmic Optimization for Prompt Attention}

\begin{table}
    \centering
    \footnotesize
    \setlength{\abovecaptionskip}{2pt}
    \caption{Evaluating the effectiveness of manually designed versus algorithmically optimized prompt attention in \system. \system-M denotes \system based on manually designed prompt attention.}
    \setlength{\tabcolsep}{4mm}
    \begin{tabular}{ccccc}
        \toprule
        \multirow{3}{*}{\textbf{Attack Method}} & \multicolumn{3}{c}{\textbf{Metric}} \\ 
        \cmidrule(l){2-4} 
        & \begin{tabular}[c]{@{}c@{}}{Semantic} \\{Similarity} \end{tabular} &
        \begin{tabular}[c]{@{}c@{}}{Syntactic} \\{Similarity} \end{tabular} &
        \begin{tabular}[c]{@{}c@{}}{Structural} \\{Similarity} \end{tabular} \\ 
        \midrule
        {\system-M} & 0.71 & 0.69 & 0.89 \\ 
        {\system} & \textbf{0.80} & \textbf{0.88} & \textbf{0.91} \\ 
        \bottomrule
    \end{tabular}
    \label{tab:prompt_attention_compare}
\end{table}
We evaluate the feasibility of manual construction versus algorithmic optimization for prompt attention. The manual method, leveraging prior knowledge, offers cost benefits. We incorporate key factors like theme, argument, structure, style, tone, vocabulary, sentence patterns, purpose, audience, and background~\cite{bernard1998text}, guiding the generative models to focus on these aspects to capture the detailed intent of the target prompts. Table~\ref{tab:prompt_attention_compare} shows that \system-M performs worse than \system. Notably, in semantic similarity, \system-M scores on average $12.7\%$ lower than \system. For syntactic similarity, \system-M scores on average $27.5\%$ lower than \system. This result suggests that manually generating prompt attention is less effective than algorithmic optimization.

\section{Transferability}

\begin{table}[]
\centering
\footnotesize 
\caption{Comparative average transferability performance of \system.}
\label{tab:transferability}
\begin{tabular}{@{}C{1.2cm}C{1.7cm}C{2cm}C{2cm}@{}}
\toprule
\multirow{4}{*}{\textbf{Metric}} & \multirow{4}{*}{\begin{tabular}[c]{@{}c@{}}{\textbf{Generative}} \\ {\textbf{Model}}\end{tabular}} & \multicolumn{2}{c}{\textbf{Target Prompt}} \\ \cmidrule(l){3-4} 
 &  & GPT-3.5 Based Prompt & GPT-4 Based Prompt \\
\midrule
\multirow{2}{*}{\begin{tabular}[c]{@{}c@{}}{Semantic} \\ {Similarity}\end{tabular}} 
 & GPT-3.5 & \textbf{0.79} & 0.75 \\
 & GPT-4 & 0.73 & \textbf{0.80} \\
\midrule
\multirow{2}{*}{\begin{tabular}[c]{@{}c@{}}{Syntactic} \\ {Similarity}\end{tabular}} 
 & GPT-3.5 & 0.88 & 0.86 \\
 & GPT-4 & \textbf{0.89} & \textbf{0.90} \\
\midrule
\multirow{2}{*}{\begin{tabular}[c]{@{}c@{}}{Structural} \\ {Similarity}\end{tabular}} 
 & GPT-3.5 & \textbf{0.91} & 0.89 \\
 & GPT-4 & 0.90 & \textbf{0.92} \\
\bottomrule
\end{tabular}
\end{table}

Consider that some real-world prompt services may not disclose their target LLMs. This lack of model disclosure highlights the importance of evaluating the transferability of attack methods to unknown target LLMs. Based on the current state of commercial prompt services, we demonstrate the transferability of \system between the popular target LLMs, GPT-3.5 and GPT-4. As shown in Table~\ref{tab:transferability}, when \system uses GPT-3.5 as the generative model, the semantic similarity applied to GPT-4 based prompts decreases by $6.3\%$, syntactic similarity decreases by $4.4\%$, and structural similarity decreases by $3.3\%$. Conversely, when GPT-4 is used as the generative model, the semantic similarity applied to GPT-3.5 prompts decreases by $7.6\%$, while syntactic and structural similarities remain almost unchanged.
Overall, the transfer performance loss of \system between GPT-3.5 and GPT-4 is under $10\%$, indicating good transferability across GPT versions.

\section{Robustness of Semantic Similarity under Different Embeddings}
\label{appendix: robustness_semantic_sim}
To assess whether the semantic similarity scores in our functional consistency metric are sensitive to the choice of embedding model, we conduct an embedding-substitution experiment by randomly selecting 200 prompts from PromptBase. We replace the original sentence transformer (specifically, a Sentence-BERT encoder) with four widely used alternatives: OpenAI’s \texttt{text-embedding-3-small}~\cite{text-embedding-3-small}, Microsoft’s \texttt{e5-base-v2}~\cite{e5-base-v2}, BAAI’s \texttt{bge-base-en-v1.5}~\cite{bge-base-en-v1.5}, and Google’s \texttt{gtr-t5-base}~\cite{gtr-t5-base}. For each embedding model, we compute semantic similarity scores between outputs generated by the stolen and target prompts, and rank the samples accordingly. We evaluate the agreement between rankings using Kendall’s Tau~\cite{kendall1938new}, a rank correlation coefficient where higher values (closer to 1) indicate stronger consistency.

\begin{table}
\footnotesize
\centering
\caption{Kendall’s Tau measuring ranking consistency between the original sentence transformer and other embedding models.}
\label{table:embedding_tau}
\begin{tabular}{ccccc}
\toprule
\multirow{3}{*}{\textbf{Metric}} & \multicolumn{4}{c}{\textbf{Embedding Model}} \\
\cmidrule(lr){2-5}
 & \makecell[c]{text-\\embedding-3-small} 
 & \makecell[c]{e5-\\base-v2} 
 & \makecell[c]{bge-base-\\en-v1.5} 
 & \makecell[c]{gtr-t5-\\base} \\
\midrule
Kendall’s Tau & 0.72 & 0.71 & 0.72 & 0.70 \\
\bottomrule
\end{tabular}
\end{table}

As shown in Table~\ref{table:embedding_tau}, all Kendall’s Tau values exceed $0.70$ across different embedding models. This indicates that while the absolute similarity scores vary, the relative ranking trends remain stable. These results confirm that the semantic similarity scores used in our functional consistency metric are robust to the choice of embedding model.

\section{Performance under Varying Numbers of Input-Output Pairs}

\begin{figure}
	\setlength{\abovecaptionskip}{1pt}
	\captionsetup[subfigure]{justification=centering}
	\centering
	\begin{subfigure}{0.325\linewidth}
		\includegraphics[width=\textwidth]{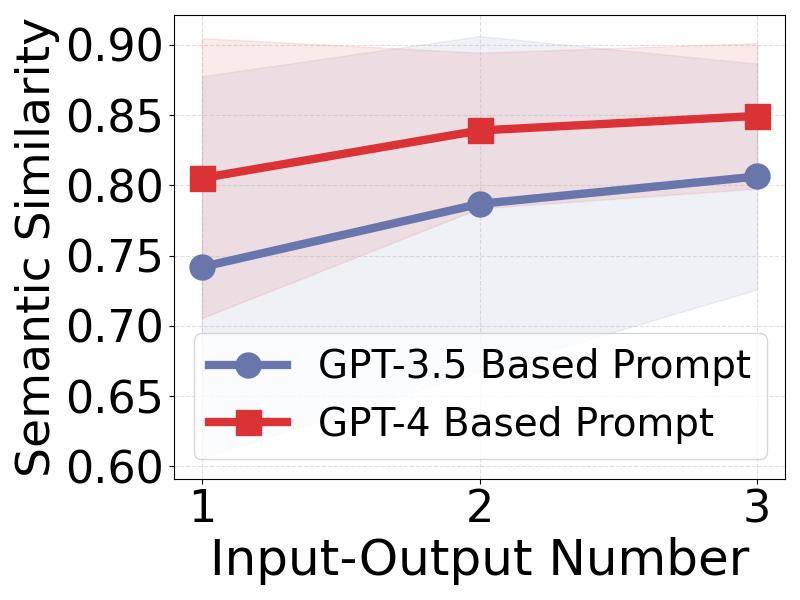}
		\caption{\scriptsize{Semantic Similarity}}\label{fig:multi_io_semantic}
	\end{subfigure}
	\begin{subfigure}{0.325\linewidth}
		\includegraphics[width=\textwidth]{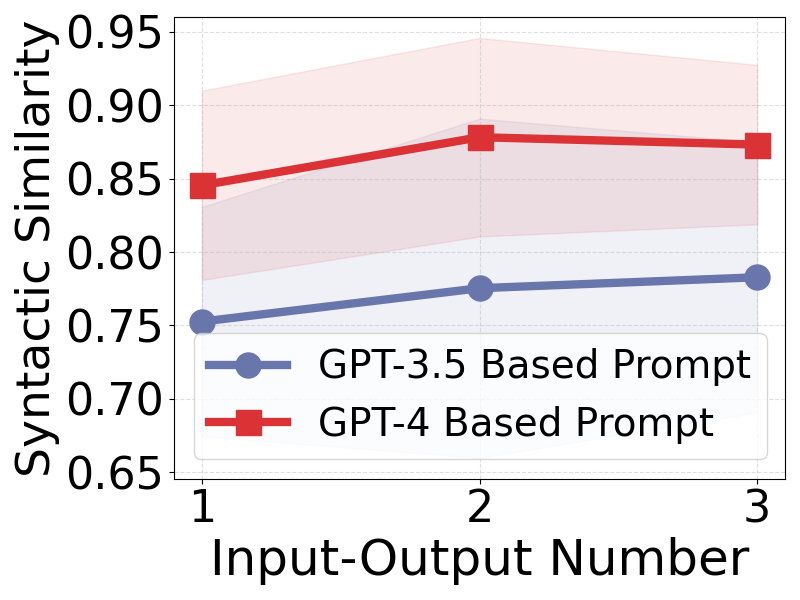}
		\caption{\scriptsize{Syntactic Similarity}}\label{fig:multi_io_syntactic}
	\end{subfigure}
	\begin{subfigure}{0.325\linewidth}
		\includegraphics[width=\textwidth]{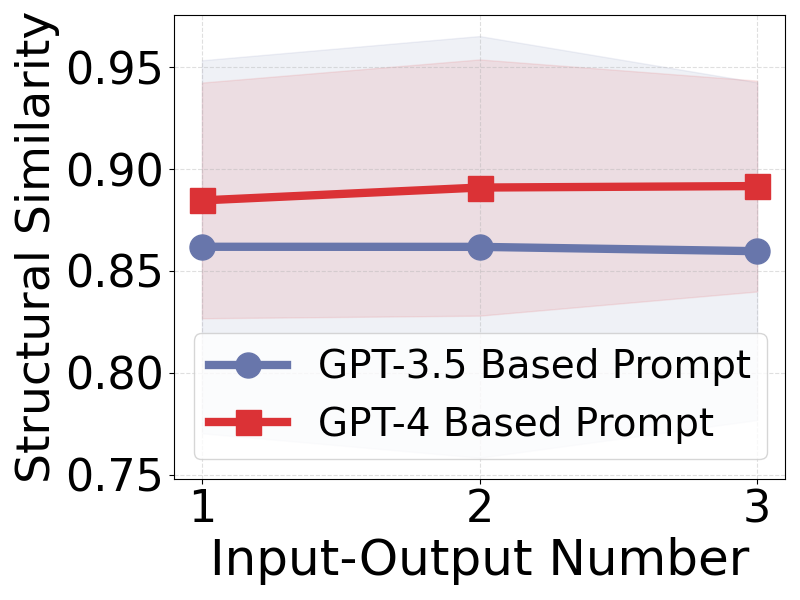}
		\caption{\scriptsize{Structural Similarity}}\label{fig:multi_io_structural}
	\end{subfigure}
  \caption{Attack performance of \system under different numbers of input-output pairs.}
  \label{fig:multi_io}
\end{figure}

To evaluate whether \system's performance benefits from access to more input-output pairs, we vary the number of pairs per target prompt from $1$ to $3$. Figure~\ref{fig:multi_io} shows the resulting functional consistency scores. We observe consistent performance improvements as the number of pairs increases, especially from $1$ to $2$. The gains are most pronounced in semantic and syntactic similarity, while structural similarity remains relatively stable. These results suggest that \system can more accurately capture the target prompt’s intent when given multiple input-output pairs.

\section{Summary of Interactions with Vendors and Developers}
As of now, we have received the response from OpenAI, who expressed gratitude for our responsible disclosure and acknowledged the value of our research. We also contacted the developers of the evaluated prompts and GPTs, sharing the identified risks along with proposed mitigation strategies. Most developers responded positively and expressed interest in collaborating on long-term defense solutions. Given the complexity of prompt protection, we also recommended interim measures, such as temporarily removing high-risk prompts or GPTs. Subsequently, we observed that several reported high-risk prompts were removed from PromptBase and that some GPTs' system prompts and protective instructions (e.g., ``Math'') have been modified. Due to confidentiality constraints, we are unable to verify the specific changes made.

\section{Future Work}
In this paper, we focus on prompt stealing attacks involving textual inputs and outputs. An important direction for future research is to explore such attacks in the context of LLM-based agents. Although \system is not designed for any specific category, its performance on less common categories remains to be further investigated. Another promising direction is to develop defense strategies that mitigate prompt stealing risks without compromising the usability and performance of prompts.

\end{document}